\documentclass[10pt]{iopart}

\usepackage{gensymb}
\usepackage{amssymb}
\usepackage{xcolor}
\usepackage{cite}
\usepackage{graphicx}
\usepackage{multirow}
\usepackage{xfrac}
\usepackage{hyperref}
\usepackage[utf8]{inputenc}

\DeclareUnicodeCharacter{2061}{}

\usepackage{iopams}  
\expandafter\let\csname equation*\endcsname\relax

\expandafter\let\csname endequation*\endcsname\relax

\usepackage{amsmath}

\begin{document}

\title[]{Neutron Diffraction Evidence for Local Spin Canting, Weak Jahn-Teller Distortion, and Magnetic Compensation in Ti$_{1-x}$Mn$_{x}$Co\textsubscript{2}O\textsubscript{4} Spinel}

\author{P. Pramanik\textsuperscript{1}, D. C. Joshi\textsuperscript{1,4}, M. Reehuis \textsuperscript{2}, A. Hoser\textsuperscript{2}, J.-U. Hoffmann\textsuperscript{2}, R. S. Manna\textsuperscript{3}, T. Sarkar\textsuperscript{4} and S. Thota\textsuperscript{1}}

\address{\textsuperscript{1}Department of Physics, Indian Institute of Technology Guwahati, Guwahati 781039, India\\
\textsuperscript{2}Helmholtz-Zentrum Berlin f$\ddot{u}$r Materialien und Energie, Hahn-Meitner-Platz 1, D-14109 Berlin, Germany\\
\textsuperscript{3}Department of Physics, Indian Institute of Technology Tirupati, Tirupati 517506, AP, India\\
\textsuperscript{4}Department of Engineering Sciences, Uppsala University, Box 534, SE-751 21 Uppsala, Sweden}

\vspace{10pt}

\begin{indented}
\item[]
\end{indented}

\begin{abstract}

A systematic study using neutron diffraction and magnetic susceptibility are reported on Mn substituted ferrimagnetic inverse spinel Ti$_{1-x}$Mn$_{x}$Co\textsubscript{2}O\textsubscript{4} in the temperature interval 1.6 K $\leq$ \emph{T} $\leq$ 300 K. Our neutron diffraction study reveals cooperative distortions of the \emph{T}O\textsubscript{6} octahedra in Ti$_{1-x}$Mn$_{x}$Co\textsubscript{2}O\textsubscript{4} system for all the Jahn-Teller active ions \emph{T} = Mn\textsuperscript{3+}, Ti\textsuperscript{3+} and Co\textsuperscript{3+}, having the electronic configurations
3\emph{d}\textsuperscript{1}, 3\emph{d}\textsuperscript{4} and 3\emph{d}\textsuperscript{6}, respectively which are confirmed by the X-ray photoelectron spectroscopy. Two specific compositions (\emph{x} = 0.2 and 0.4) have been chosen in this study because these two systems show unique features such as; (\emph{i}) Noncollinear Yafet-Kittel type magnetic ordering, and (\emph{ii}) Weak tetragonal distortion with \emph{c/a} \textless{} 1, in which the apical bond length \emph{d\textsubscript{c}}(\emph{T\textsubscript{B}}-O) is longer than the equatorial bond length \emph{d\textsubscript{ab}}(\emph{T\textsubscript{B}}-O) due to the splitting of the \emph{e\textsubscript{g}} level of
Mn\textsuperscript{3+} ions into $d_{x^2-y^2}$ and $d_{z^2}$. For the composition \emph{x} = 0.4, the distortion in the \emph{T\textsubscript{B}}O\textsubscript{6} octahedra is stronger as compared to \emph{x} = 0.2 because of the higher content of trivalent Mn. Ferrimagnetic ordering in Ti\textsubscript{0.6}Mn\textsubscript{0.4}Co\textsubscript{2}O\textsubscript{4} and Ti\textsubscript{0.8}Mn\textsubscript{0.2}Co\textsubscript{2}O\textsubscript{4} sets in at 110.3 and 78.2 K, respectively due to the presence of unequal magnetic moments of cations, where Ti\textsuperscript{3+},
Mn\textsuperscript{3+}, and Co\textsuperscript{3+} occupying the octahedral, whereas, Co\textsuperscript{2+} sits in the tetrahedral site. For both compounds an additional weak antiferromagnetic component could be observed lying perpendicular to the ferrimagnetic component. The analysis of static and dynamic magnetic susceptibilities combined with the heat-capacity data reveals a
magnetic compensation phenomenon (MCP) at \emph{T}\textsubscript{COMP} = 25.4 K in
Ti\textsubscript{0.8}Mn\textsubscript{0.2}Co\textsubscript{2}O\textsubscript{4} and a reentrant spin-glass behaviour in Ti\textsubscript{0.6}Mn\textsubscript{0.4}Co\textsubscript{2}O\textsubscript{4}
with a freezing temperature $\sim$110.1 K. The MCP in this compound is characterized by sign reversal of magnetization and bipolar exchange bias effect below \emph{T}\textsubscript{COMP} with its magnitude depending on the direction of external magnetic field and the cooling protocol.

\end{abstract}

\ioptwocol

\section{Introduction}
Competing magnetic exchange interactions between the cations in spinel oxides \emph{AB} \textsubscript{2}O\textsubscript{4} play a major role in deciding their magnetic and crystal structure along with various other physical properties \cite{nii2013interplay,lee2015diluted,tanaka2009dilution,ohtani2010orbital,reehuis2015competing,guillou2011magnetic,hoshi2007magnetic,srinivasan1983magnetic}. The contribution of orbital degeneracy under very high crystal-field (\textgreater{}2.36 eV), Hund's exchange couplings ($\sim$0.7 eV) and spin-orbit interactions (5.5 eV)leads to some interesting properties like spin-liquid state, magnetothermal, anisotropic magnetoelastic, magnetodielectric, and
magneto-structural distortion driven by cooperative Jahn-Teller effect \cite{roth1964magnetic,lee2015diluted,reehuis2015competing,guillou2011magnetic,tomiyasu2011molecular,suzuki2009magnetodielectric,melot2009magnetic,kemei2012evolution,tackett2007magnetodielectric,suzuki2008magnetodielectric}. In particular, giant magnetic anisotropy (650 kOe) and atomic displacements ($\sim$1/4 \AA), negative-magnetization with compensation phenomena, zero-field cooled bipolar exchange-bias effect and reentrant spin-glass behavior in few Mn-, Co- and Cr-based spinels has drawn immense attention \cite{hirai2013giant,kim2012giant,padam2013magnetic,hubsch1982semi,Thota2017,zhang2015site}. Exploring the tunability of such exotic properties may lead to potential applications
in magnetic read/write heads, spin-valves and other switching devices \cite{kulikowski1980properties,carey2002spin,jiang2014resistive,hu2012opportunity}.

Precise understanding of the cationic distribution using the neutron diffraction studies in diluted magnetic spinels is a challenging issue. For example, earlier studies related to the dilution of
Co\textsubscript{3}O\textsubscript{4} at octahedral sites with Mg (ionic radius \emph{2r}\textsubscript{IV} = 0.57 \AA) reveals that there is a limit to the occupancy of the Mg\textsuperscript{2+} ions at tetrahedral sites \cite{krezhov1992cationic}. Such dilution (with tetravalent nonmagnetic elements like Sn\textsuperscript{4+}, Ge\textsuperscript{4+} or
Ti\textsuperscript{4+}) play a major role in the disruption of antiferromagnetic ordering in Co\textsubscript{3}O\textsubscript{4} and results in a short-range ordering with reentrant spin-glass (RSG) behavior \cite{hubsch1982semi,thota2013co,mandrus1999spin,Thota2017}. Coexistence of ferrimagnetic behaviour with RSG, negative magnetization below the compensation temperature and giant
asymmetry in the hysteresis loops are some of the important features noticed in such diluted inverse spinels \cite{hubsch1982semi,thota2013co,mandrus1999spin,Thota2017,li2018negative}. Nevertheless, diluting the octahedrally coordinated Co\textsuperscript{3+} ions with magnetic element Mn\textsuperscript{3+} leads to the formation of inverse spinel MnCo\textsubscript{2}O\textsubscript{4} which has tremendous applications in the renewable energy sector, such as, solid-oxide fuel cells, lithium-ion batteries, catalysis and magnetoelectronics \cite{fu2014one,molin2016low,yang2005thermal}. Here it is important to note that the Co\textsuperscript{3+} ions at the \emph{B} site carries no magnetic moment in Co\textsubscript{3}O\textsubscript{4} \cite{roth1964magnetic} and
MnCo\textsubscript{2}O\textsubscript{4} \cite{boucher1970etude} in the low-spin state due to a large crystal-field splitting of the \emph{d} orbitals. On the other hand it could be shown that the Co\textsuperscript{3+} ions in TiCo\textsubscript{2}O\textsubscript{4} carry a magnetic moment
\cite{Thota2017}. Therefore, the aim of our current research work is two-fold, firstly, to get more information about the magnitude and valence of the magnetic moments of the individual magnetic metal ions using mixed spinel oxides for various compositions of Ti$_{1-x}$Mn$_{x}$Co\textsubscript{2}O\textsubscript{4}. Secondly, majority of the reports available in the literature on both end members  TiCo\textsubscript{2}O\textsubscript{4} and MnCo\textsubscript{2}O\textsubscript{4} are primarily focused on the synthesis and characterization of the porous nanostructures with special emphasis on their catalytic and electrochemical properties, but little literature is available on the low-temperature neutron diffraction studies with special emphasis on their magnetic ground states and
cationic distribution \cite{roth1964magnetic,granroth2004long,mondal2015mesoporous,li2014one,yuvaraj2016synthesis,arrebola2008improving}. In particular, a detailed study of different compositions of bulk grain sized Ti$_{1-x}$Mn$_{x}$Co\textsubscript{2}O\textsubscript{4} systems is still lacking in the literature. Moreover, these solid solutions are interesting mainly because of the unique magnetic characteristics of both the end compounds TiCo\textsubscript{2}O\textsubscript{4} and MnCo\textsubscript{2}O\textsubscript{4} reported earlier. Some of the intriguing properties of these compounds being: (\emph{i}) co-existence
of longitudinal ferrimagnetic order and transverse spin-glass state below freezing point resulting from the dilutants on the \emph{B} sites, (\emph{ii}) negative magnetization below the compensation point and bipolar exchange bias under ZFC condition \cite{hubsch1982semi,Thota2017,srivastava1987spin}. Moreover, the inverse spinel MnCo\textsubscript{2}O\textsubscript{4} exhibits only ferrimagnetic behavior, without any of the above mentioned characteristics. Since both the end compounds exhibits very different magnetic ground state it is worth to investigate the magnetic and crystal structures of the intermediate compositions of these two end compounds. Another motivating factor in investigating this system is to probe the magnitude of permanent magnetic moment on the trivalent \emph{B} site Co ions which usually do not exhibit any magnetic moment due to the large crystal-field splitting ($\geq$19000 cm\textsuperscript{-1})of the 3\emph{d} orbitals by the octahedral cubic field \cite{roth1964magnetic}. Therefore, motivated by the wide range of applications and unique
magnetic characteristics of MnCo\textsubscript{2}O\textsubscript{4} and TiCo\textsubscript{2}O\textsubscript{4} compounds, an attempt has been made in the present work to investigate a detailed crystal structure and magnetic ordering of Ti$_{1-x}$Mn$_{x}$Co\textsubscript{2}O\textsubscript{4}
polycrystalline samples using neutron powder diffraction and dc- and ac-magnetization measurements. Among various compositions (0 $\leq$ \emph{x} $\leq$ 1) that we prepared, two specific compositions
Ti\textsubscript{0.8}Mn\textsubscript{0.2}Co\textsubscript{2}O\textsubscript{4} and Ti\textsubscript{0.6}Mn\textsubscript{0.4}Co\textsubscript{2}O\textsubscript{4} are chosen for a detailed study because they show the evidence for magnetic compensation effect, reentrant spin-glass behavior and exchange-bias effect.

Based on the characterization results; (\emph{i}) we examine the following statistical distribution of cationic occupancy (Co\textsuperscript{2+})\emph{\textsubscript{A}}(Ti\textsubscript{0.8}\textsuperscript{3+}Mn\textsubscript{0.2}\textsuperscript{3+}Co\textsuperscript{3+})\emph{\textsubscript{B}}O\textsubscript{4} and\\(Co\textsuperscript{2+})\emph{\textsubscript{A}}(Ti\textsubscript{0.6}\textsuperscript{3+}Mn\textsubscript{0.4}\textsuperscript{3+}Co\textsuperscript{3+})\emph{\textsubscript{B}}O\textsubscript{4}, (\emph{ii}) large bipolar exchange-bias fields (3.67 kOe at \emph{T} = 15 K), (\emph{iii}) negative magnetization ($-$0.67 emu/g at \emph{T} = 5 K), and (\emph{iv}) magnetic compensation phenomena (\emph{T}\textsubscript{COMP} $\sim$ 25.4 K) for \emph{x} =
0.2, and (\emph{v}) the determination of the ordered magnetic moment values and valencies of the individual metal ions using spinel mixed oxides.

\section{Experimental Details}
Polycrystalline samples of Mn doped TiCo\textsubscript{2}O\textsubscript{4} were fabricated using the
standard solid-state reaction method using stoichiometric amounts of the precursors TiO\textsubscript{2}, Mn\textsubscript{2}O\textsubscript{3}, and Co\textsubscript{3}O\textsubscript{4}. These mixed oxides were grounded in an agate mortar for 6 hours and pelletized using a hydraulic
press with 50 kN pressure followed by sintering at 1100\textsuperscript{o}C for 8 hours duration with 4\textsuperscript{o}C per minute heating and cooling rates. Structural characterization and phase purity of the sintered pellets were investigated by X-ray diffraction (XRD) measurements, performed using a Rigaku X-ray diffractometer (model: TRAX III) with Cu-K$\alpha$ radiation (\emph{$\lambda$} = 1.54056 \AA). The XRD patterns of Ti$_{1-x}$Mn$_{x}$Co\textsubscript{2}O\textsubscript{4} were very similar to the end member TiCo\textsubscript{2}O\textsubscript{4} {[}space group $Fd\overline{3}m$ (No. 227){]}. However, a slight decrease in the lattice parameter was noticed with increasing the Mn concentration (\emph{a} = 8.384 \AA, for \emph{x} $\sim$ 0.2) due to smaller ionic size of the Mn (\emph{r} $\sim$ 0.645 \AA) as compared to Ti (\emph{r} $\sim$ 0.67 \AA).

Neutron powder diffraction experiments on Ti\textsubscript{0.8}Mn\textsubscript{0.2}Co\textsubscript{2}O\textsubscript{4} and Ti\textsubscript{0.6}Mn\textsubscript{0.4}Co\textsubscript{2}O\textsubscript{4}
were carried out on the instruments E2, E6, and E9 at the BER II reactor of the Helmholtz-Zentrum Berlin. The instrument E9 uses a Ge monochromator selecting the neutron wavelength $\lambda$ = 1.3083 \AA, while the instruments E2 and E6 use a pyrolytic graphite (PG) monochromator selecting the neutron wavelengths $\lambda$ = 2.379 and 2.423 \AA, respectively. On these instruments powder patterns were recorded between the following diffraction angles: between 14.7 and 90.2$\degree$ (E2), 5.5 and 136.6$\degree$ (E6), and 5 and 141.8$\degree$ (E9). For both compounds Ti\textsubscript{0.8}Mn\textsubscript{0.2}Co\textsubscript{2}O\textsubscript{4} and Ti\textsubscript{0.6}Mn\textsubscript{0.4}Co\textsubscript{2}O\textsubscript{4} the crystal structure parameters at 3 K were investigated from data sets collected on E9. An additional neutron diffraction pattern from Ti\textsubscript{0.6}Mn\textsubscript{0.4}Co\textsubscript{2}O\textsubscript{4} sample was collected for sake of comparison at 295 K. In order to investigate in detail the magnetic structure neutron powder diffraction patterns were collected at \emph{T} = 2 K on the instrument E2 with high counting statistics (24 h/pattern)
using a 15-min collimation to improve the instrumental resolution. In the paramagnetic range a second powder pattern of Ti\textsubscript{0.8}Mn\textsubscript{0.2}Co\textsubscript{2}O\textsubscript{4}
and Ti\textsubscript{0.6}Mn\textsubscript{0.4}Co\textsubscript{2}O\textsubscript{4} was collected at 90 and 142 K, respectively.

The temperature dependent behaviour of the magnetic ordering of both spinels has been investigated on the instrument E6. The refinements of crystal and magnetic structures were carried out with the
\emph{FullProf} program \cite{RODRIGUEZCARVAJAL199355}. The nuclear scattering lengths \emph{b}(O) = 5.805 fm, \emph{b}(Ti) = $-$ 3.30 fm, \emph{b}(Mn) = $-$ 3.73 fm, and \emph{b}(Co ) = 2.50 fm, were used \cite{sears1995international}. The magnetic form factors of the Ti\textsuperscript{3+}, Mn\textsuperscript{2+}, Co\textsuperscript{2+} and Co\textsuperscript{3+} ions were taken from
Ref. \cite{brown1995international}.

For the electronic structure and elemental analysis, we performed X-ray photoelectron spectroscopy (XPS) measurements using Al-K$\alpha$ lab source. All the data was recorded using an Omicron hemispherical analyzer. After subtracting the Tougard background, all the XPS core-level (CL) data
were fitted with mixed Lorentzian--Gaussian profile. The superconducting quantum interference device (SQUID) based magnetometer (MPMS: magnetic property measurements system) with working temperature range 4-300 K and magnetic field (\emph{H}) up to $\pm$50 kOe was used for the magnetization measurements. Temperature dependence of specific heat \emph{C\textsubscript{P}}(\emph{T}) was recorded by means of a
physical-property-measurement-systems (PPMS) from Quantum design.

\section{\label{sec:levelIII.} RESULTS AND DISCUSSION}

\subsection{\label{sec:level1}Crystal structure:}
The inverse spinel Co\textsubscript{2}TiO\textsubscript{4} was found to crystallize in the cubic space group $Fd\overline{3}m$ (No. 227) \cite{Thota2017}. Usually in normal spinels with the general formula \emph{AB}\textsubscript{2}O\textsubscript{4} the \emph{A}\textsuperscript{2+} ions are located on the tetrahedral site (\emph{A} site), while the \emph{B}\textsuperscript{3+} ions are located
on the octahedral site (\emph{B} site). In TiCo\textsubscript{2}O\textsubscript{4} the cobalt ions have the valence 2+ and 3+, and therefore it can be expected that Co\textsuperscript{2+} at the \emph{A} site (labelled as Co\emph{\textsubscript{A}}) is located at the Wyckoff position 8\emph{b}($\sfrac{3}{8}$,$\sfrac{3}{8}$,$\sfrac{3}{8}$), while Co\textsuperscript{3+} occupy one half of the \emph{B} site (labelled as Co\emph{\textsubscript{B}}) located at 16\emph{c}(0,0,0). Consequently, the titanium ions are expected to have the valence 3+ and occupy the other half of the \emph{B} site. Further the oxygen atoms occupy the position 32\emph{e}(\emph{x},\emph{x},\emph{x}). The formula of this inverse spinel can be given as Co(Co\textsubscript{0.5}Ti\textsubscript{0.5})\textsubscript{2}O\textsubscript{4}.
Accordingly, we have refined the crystal structure of Ti\textsubscript{0.8}Mn\textsubscript{0.2}Co\textsubscript{2}O\textsubscript{4} and Ti\textsubscript{0.6}Mn\textsubscript{0.4}Co\textsubscript{2}O\textsubscript{4} {[}or in detail Co(Ti\textsubscript{0.4}Mn\textsubscript{0.1}Co\textsubscript{0.5})\textsubscript{2}O\textsubscript{4} and Co(Ti\textsubscript{0.3}Mn\textsubscript{0.2}Co\textsubscript{0.5})\textsubscript{2}O\textsubscript{4}{]} in the cubic space group . In order to determine the correct composition, we have refined the site occupancies of both compounds from
data sets collected on E9. Due to the fact that manganese and titanium have negative neutron scattering lengths it was possible to determine the occupancies with good accuracy. In the first step the occupancies of the Co\textsuperscript{2+} ions at the \emph{A} site and the oxygen atoms were refined, while the occupancies of the \emph{B}-site atoms were fixed. This could improve the fit with the consequence that the obtained occupancies of both Co\textsuperscript{2+} and O were found to be too small. On the other hand, it is important to note that the ratio \emph{occ}(O)/\emph{occ}(Co\textsuperscript{2+}) reached a value 4.11(3) which is slightly larger than the ideal value 4. This indicates that the sample does not show an oxygen deficiency. In a further refinement we assumed that Mn atoms are also located at the \emph{A} site. It was found that only a negligible amount of 1.1(5) \% of manganese could be located at this position. This shows that the \emph{A} site is practically fully occupied with Co\textsuperscript{2+}. This was also found earlier for the compound MnCo\textsubscript{2}O\textsubscript{4}\cite{joy2000unusual}. In the next refinement the occupancies of
Co\textsuperscript{2+} and O were fixed. Here one finds a reduced scattering power at the \emph{B} site which indicates that the manganese and titanium content at the \emph{B} site is higher than that of cobalt. The loss of cobalt can be explained by the fact that both samples contain minor impurities of Co\textsubscript{3}O\textsubscript{4}. On the basis of accurate weighting of the components we finally used the constraint for the \emph{B} site (1 + \emph{x}) {[}\emph{occ}(Ti) + \emph{occ}(Mn){]} + (1$-$\emph{x}) \emph{occ}(Co) = 1, where \emph{occ} is the nominal content at the \emph{B} site. Finally, from the data sets collected for \emph{x} = 0.4 at T = 3 and 295 K we found the chemical compositions are (Co\textsuperscript{2+})\emph{\textsubscript{A}}{[}Ti\textsubscript{0.629(5)}\textsuperscript{3+}Mn\textsubscript{0.420(5)}\textsuperscript{3+}Co\textsubscript{0.951(5)}\textsuperscript{3+}{]}\emph{\textsubscript{B}}O\textsubscript{4} and (Co\textsuperscript{2+})\emph{\textsubscript{A}}{[}Ti\textsubscript{0.633(5)}\textsuperscript{3+}Mn\textsubscript{0.422(5)}\textsuperscript{3+}Co\textsubscript{0.945(5)}\textsuperscript{3+}{]}\emph{\textsubscript{B}}O\textsubscript{4}, respectively. For the other sample with \emph{x} = 0.2 the chemical composition was found to be (Co\textsuperscript{2+})\emph{\textsubscript{A}}{[}Ti\textsubscript{0.841(5)}\textsuperscript{3+}Mn\textsubscript{0.210(5)}\textsuperscript{3+}Co\textsubscript{0.949(5)}\textsuperscript{3+}{]}\emph{\textsubscript{B}}O\textsubscript{4}.

\begin{figure}[t]
\includegraphics[trim=0.7cm 0.6cm 5.3cm 0.8cm, clip=true,scale=1.4]{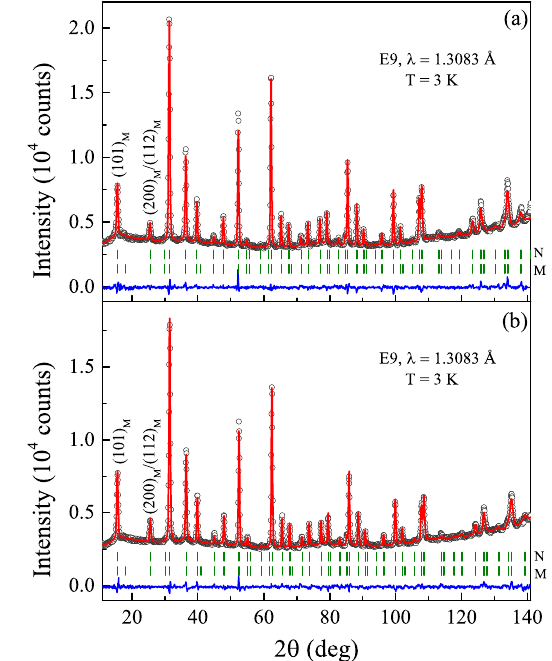}
\caption{Rietveld refinements of the neutron powder diffraction data of (\emph{a}) Ti\textsubscript{0.8}Mn\textsubscript{0.2}Co\textsubscript{2}O\textsubscript{4} and (\emph{b}) Ti\textsubscript{0.6}Mn\textsubscript{0.4}Co\textsubscript{2}O\textsubscript{4} collected on E9 (wavelength, $\lambda$ = 1.3083 \AA) at 3 K. The crystal structure was refined in the tetragonal space group \emph{I}4\textsubscript{1}/\emph{amd}. The calculated patterns (red) are compared with the observed one (black circles). In the lower part of each diagram the difference pattern (blue) as well as the positions of the nuclear (N) and magnetic (M) reflections of are shown. A strongest magnetic contribution is observed for the reflections 101 and 200/112 (see also Fig. 3).}
\label{fig:Fig1}
\end{figure}

\begin{figure}[t]
\includegraphics[trim=2cm 6.1cm 4.8cm 4.8cm, clip=true,scale=0.49]{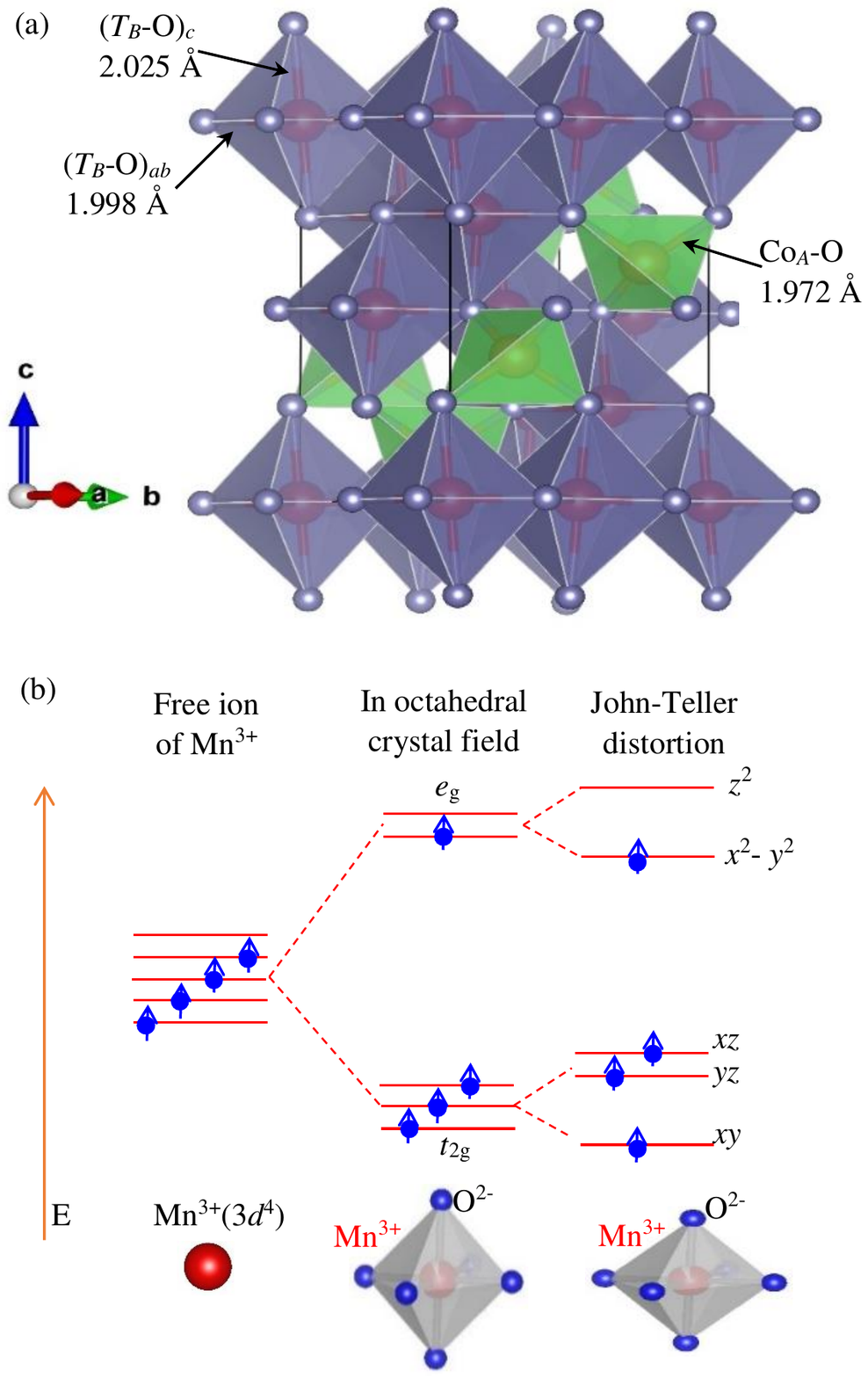}
\caption{(a) Crystal structure of tetragonally distorted Ti$_{1-x}$Mn$_{x}$Co\textsubscript{2}O\textsubscript{4} unit cell visualized by using VESTA. The tetrahedral (green) positions are occupied by Co\textsuperscript{2+}, whereas  octahedral (lightSlateBlue) sites are occupied by Mn\textsuperscript{3+}, Ti\textsuperscript{3+}, and Co\textsuperscript{3+}. (b) Schematic illustrations of Mn\textsuperscript{3+}(3\emph{d}\textsuperscript{4}) degenerate orbital energy levels. In the larger crystal field of octahedral symmetry, the degeneracy of these energy levels are lifted into triply \emph{t}\textsubscript{2g} and doubly \emph{e}\textsubscript{g} orbitals. The Mn\textsuperscript{3+}O\textsubscript{6} octahedra became distorted due to Jahn$-$Teller effect.}
\label{fig:Fig2}
\end{figure}

Despite the fact that we could not observe a peak broadening of particular reflections, we also tried to carry out the refinements of the crystal structure in the next lower symmetric space group
\emph{I}4\textsubscript{1}/\emph{amd} (No. 141, cell choice 2) with the cell dimensions \emph{a}\textsubscript{t} $\times$ \emph{b}\textsubscript{t} $\times$ \emph{c}\textsubscript{t} = \emph{a}\textsubscript{c}/$\sqrt{2}$ $\times$ \emph{b}\textsubscript{c}/$\sqrt{2}$ $\times$ \emph{c}\textsubscript{c}. In this setting the Co\textsuperscript{2+} ions at the \emph{A} site (labelled
as Co\emph{\textsubscript{A}}) are located at the Wyckoff position 4\emph{b}(0,1/4,3/8), while at the \emph{B} site the Ti\textsuperscript{3+}, Mn\textsuperscript{3+} ions, and the other half of the Co (here as Co\textsuperscript{3+} and labelled as Co\emph{\textsubscript{B}}) are located at 8\emph{c}(0,0,0). The oxygen atoms occupy the position 16\emph{h}(0,\emph{y},\emph{z}), where \emph{y} and \emph{z} approximately have the values 0.5 and 0.25, respectively. However, we found from the Rietveld
refinements of Ti\textsubscript{0.6}Mn\textsubscript{0.4}Co\textsubscript{2}O\textsubscript{4} system \emph{c}/\emph{a}$\sqrt{2}$ ratios to be 0.9988(3) and 0.9995(3) at 3 and 295 K, respectively. Similarly, \emph{c}/\emph{a}$\sqrt{2}$ ratio for Ti\textsubscript{0.8}Mn\textsubscript{0.2}Co\textsubscript{2}O\textsubscript{4} at 3 K was found to be 0.9988(3). These values indicate the presence of a weak tetragonal distortion at low temperature in the magnetically ordered region. The results of the Rietveld refinements are given in table 1 and figure 1. For the Co\textsuperscript{2+} ions at the \emph{A} site, which have the 3\emph{d}\textsuperscript{7} configuration, cooperative distortions of the CoO\textsubscript{4} tetrahedra through the Jahn-Teller effect are absent. This is not the case for the \emph{T}O\textsubscript{6} octahedra, which contain the Jahn-Teller active ions
Ti\textsuperscript{3+}, Mn\textsuperscript{3+} and Co\textsuperscript{3+}, having the 3\emph{d}\textsuperscript{1}, 3\emph{d}\textsuperscript{4} and 3\emph{d}\textsuperscript{6} configurations, respectively. For the Mn\textsuperscript{3+}, Ti\textsuperscript{3+} and Co\textsuperscript{3+} ions electronic energy can be gained if the 2\emph{t\textsubscript{g}} levels split into a lower \emph{d\textsubscript{xy}} level and a higher twofold generate \emph{d\textsubscript{xz}}/\emph{d\textsubscript{yz}} level (figure 2). This would result in a tetragonal distortion with a \emph{c}/\emph{a} ratio smaller than 1, and also a shrinking of the apical bond \emph{d\textsubscript{c}}(\emph{T}-O) (figure 2). But the crystal structure refinements of Ti\textsubscript{0.8}Mn\textsubscript{0.2}Co\textsubscript{2}O\textsubscript{4} showed that the apical bond length
\emph{d\textsubscript{c}}(\emph{T\textsubscript{B}}-O) = 2.020(5) \AA~is slightly longer than the equatorial one \emph{d\textsubscript{ab}}(\emph{T\textsubscript{B}}-O) = 2.016(2) \AA. This
effect is stronger in Ti\textsubscript{0.6}Mn\textsubscript{0.4}Co\textsubscript{2}O\textsubscript{4},
where the Mn content is twice than that in Ti\textsubscript{0.8}Mn\textsubscript{0.2}Co\textsubscript{2}O\textsubscript{4}, in which the bond lengths are \emph{d\textsubscript{c}}(\emph{T\textsubscript{B}}-O) = 2.025(4) \AA~and \emph{d\textsubscript{ab}}(\emph{T\textsubscript{B}}-O) = 1.998(2) \AA. Practically the same values are obtained if we assume \emph{c}/\emph{a} ratio equal to 1. Here it is interesting to note that a tetragonal distortion could not be detected in TiCo\textsubscript{2}O\textsubscript{4} \cite{Thota2017} suggesting that the Jahn-Teller activity of the Ti\textsuperscript{3+} and Co\textsuperscript{3+} ions is rather weak in this system. Therefore, we can assume that the elongation of the apical bond \emph{d\textsubscript{c}}(\emph{T}-O) can be ascribed to the Jahn-Teller activity of the Mn\textsuperscript{3+} ions having the 3\emph{d}\textsuperscript{4} configuration. In this case electronic energy can be gained if the \emph{e\textsubscript{g}} level splits into $d_{x^2-y^2}$ and $d_{z^2}$ levels, where the fourth electron occupies lower lying $d_{z^2}$ level. This electronic configuration finally leads to an elongation of the \emph{T}O\textsubscript{6} octahedra along the \emph{c} axis. Possibly the Jahn-Teller effect on the two ions act along different directions as found earlier in Ni$_{1-x}$Cu$_{x}$Cr\textsubscript{2}O\textsubscript{4} system \cite{reehuis2015competing} and leads to compensation of particular distortions.

\subsection{\label{sec:level2}Magnetic ordering:}
Figure 3 shows the neutron powder diffraction patterns of Ti\textsubscript{0.8}Mn\textsubscript{0.2}Co\textsubscript{2}O\textsubscript{4} and Ti\textsubscript{0.6}Mn\textsubscript{0.4}Co\textsubscript{2}O\textsubscript{4} collected on the instrument E2 well below and above the magnetic ordering temperature \emph{T}\textsubscript{C}. The powder patterns collected at 2 K show additional magnetic intensities due to the magnetic ordering of the Ti, Mn, and Co atoms. Due to the weak tetragonal splitting, we used the cubic setting for data analysis. For both compounds, the strongest magnetic intensity could be
observed at the position of the reflection 111 indicating a ferrimagnetic (FI) coupling between the atoms located at \emph{A} and \emph{B} sites. A much weaker magnetic intensity could be observed
at the position of the reflection 200. Therefore, the two magnetic reflections are shown in the inset of figure 3 in an enlarged view using a logarithmic scale. The reflection 200 is forbidden for the cubic space group $Fd\overline{3}m$, but also for the next-lower symmetric tetragonal space group
\emph{I}4\textsubscript{1}/\emph{amd}, where the reflections are indexes as 110 and 002. Thus the presence of magnetic intensity suggests loss of at least one of the \emph{d}-glide planes indicating the presence of an additional antiferromagnetic (AF) component. 

\begin{figure}[t]
\includegraphics[trim=0.7cm 0.5cm 0cm 0.7cm, clip=true,scale=0.8]{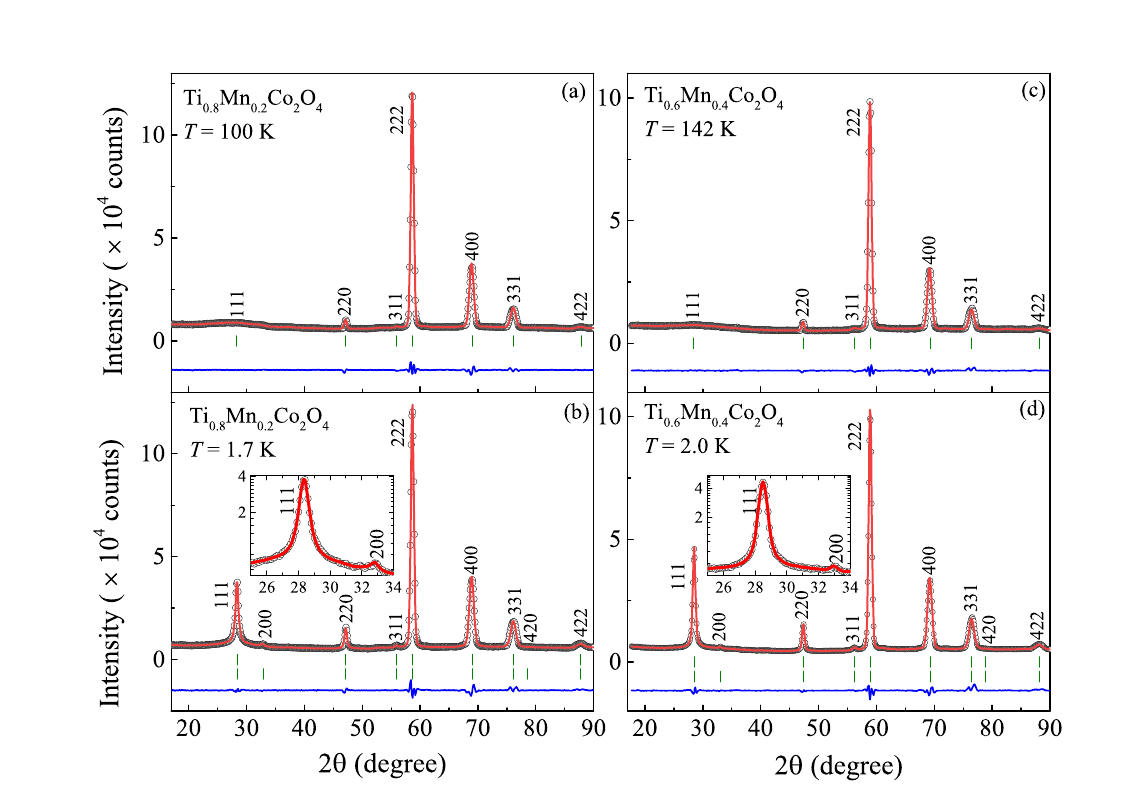}
\caption{Neutron powder patterns of Ti\textsubscript{0.8}Mn\textsubscript{0.2}Co\textsubscript{2}O\textsubscript{4} and Ti\textsubscript{0.6}Mn\textsubscript{0.4}Co\textsubscript{2}O\textsubscript{4} collected on the instrument E2 at temperatures well below and above the long-range magnetic ordering. The crystal structure was refined in the cubic space group  . The powder patterns in the lower part of the diagram show additional magnetic intensities due to the magnetic ordering of the Ti, Mn, and Co atoms. The insets present the prominent magnetic reflections (111)\textsubscript{M} and (200)\textsubscript{M} in enlarged form. Due to the weakness of the reflection (200)\textsubscript{M} we used a logarithmic scale. The calculated patterns (red) are compared with the observed ones (black circles). The difference patterns (blue) as well as the positions of the nuclear and magnetic Bragg reflections are also shown.}
\label{fig:Fig3}
\end{figure}

Using the tetragonal setting the obtained splitting of the individual reflections 110 and 002 is too weak to distinguish their magnetic contributions. Consequently, it was not possible to determine the directions of the AF and FI components. However, it was possible to determine the spin sequence of the \emph{B}-site atoms Co\emph{\textsubscript{B}}, Ti\emph{\textsubscript{B}}, and Mn\emph{\textsubscript{B}} atoms located at the positions (1) 0,0,0; (2) $\sfrac{3}{4}$,$\sfrac{1}{4}$,$\sfrac{1}{2}$; (3) $\sfrac{1}{4}$,$\sfrac{1}{2}$,$\sfrac{3}{4}$; (4) $\sfrac{1}{2}$,$\sfrac{3}{4}$,$\sfrac{1}{4}$. In general, the possible spin sequences for the \emph{B}-site atoms are +~+~+~+, +~$-$~+~$-$, +~+~$-$~$-$, and ~+~$-$~$-$~+. For the determination of the magnetic structure we fixed the moment of the ferromagnetic (F) component to be aligned parallel to the \emph{a} axis. Several tests
have shown that the best fit was achieved using the sequences +~+~$-$~$-$ and +~$-$~+~$-$ for the AF components aligned parallel to the \emph{b} and \emph{c} axes, respectively. These modes are compatible with those of the irreducible representation (\emph{irrep}) $\Gamma$\textsubscript{8} obtained from the representation analysis using the program \emph{BasIrep} of the \emph{Fullprof} suite \cite{Thota2017}. This \emph{irrep} is of dimension 3 and it appears 2 times in $\Gamma$\textsubscript{8} which means that the ferromagnetic component can be alternatively aligned parallel to \emph{b} and \emph{c}. It can be seen in figure 4 that AF coupled moments show a non collinear magnetic ordering, where in the present setting the moment points along {[}011{]} and {[}01$-$1{]}, respectively. A coexistence of AF and FI ordering was also observed in the system Ni$_{1-x}$Cu$_{x}$Cr\textsubscript{2}O\textsubscript{4} \cite{reehuis2015competing}, where the FI ordering occurs in the \emph{ab} plane and the AF component is aligned parallel to the \emph{c} axis.

Both the magnetic ions Mn\textsuperscript{3+} and Co\textsuperscript{3+} in the 3\emph{d} \textsuperscript{4} and 3\emph{d}\textsuperscript{6} configurations have four unpaired electrons in the high-spin state, while Ti\textsuperscript{3+} ions in the 3\emph{d}\textsuperscript{1} has only one electron. Therefore, the expected theoretical magnetic moment values for the high-spin state are $\mu$\textsubscript{eff} = \emph{g}\emph{S}$\mu$\textsubscript{B} = 4.0 $\mu$\textsubscript{B} for
Mn\textsuperscript{3+} and Co\textsuperscript{3+}, and $\mu$\textsubscript{eff} = \emph{g}\emph{S}$\mu$\textsubscript{B} = 1.0 $\mu$\textsubscript{B} for Ti\textsuperscript{3+}. At this point it is worth to mentioned that the total moments of the ions at the \emph{B} site can differ, especially due to the presence of spin-orbital coupling. Nevertheless, we have tried to estimate the individual moment values at the \emph{B} site using the constraint $\mu$(Mn\textsuperscript{3+}) = ~$\mu$(Co\textsuperscript{3+}) = 4$\times$$\mu$(Ti\textsuperscript{3+}). The results of the refinements are given in table 2. The obtained total moments of the Ti\textsuperscript{3+} ions vary in both compounds between 0.52 and 0.60 $\mu$\textsubscript{B}, and those of the Mn\textsuperscript{3+}/Co\textsuperscript{3+} ions between 2.09 and 2.39 $\mu$\textsubscript{B}, which are considerably smaller than the theoretical values given above. It has been mentioned above that the magnetic moments of the \emph{B} site Co\textsuperscript{3+} ions were found to be zero in Co\textsubscript{3}O\textsubscript{4} and MnCo\textsubscript{2}O\textsubscript{4} \cite{roth1964magnetic,boucher1970etude}, while the in case of TiCo\textsubscript{2}O\textsubscript{4} they are also magnetically ordered \cite{Thota2017}. In order to estimate the magnetic moments of Co\textsuperscript{3+} we used the fixed moment value 3.84 $\mu$\textsubscript{B} of Mn\textsuperscript{3+} as determined earlier for
MnCo\textsubscript{2}O\textsubscript{4} \cite{boucher1970etude}. From the refinements, where we now used the constraint $\mu$(Mn\textsuperscript{3+}) = 4$\times$$\mu$(Ti\textsuperscript{3+}), magnetic moment values of Co\textsuperscript{3+} vary between 0.75 and 0.80 for Ti\textsubscript{0.8}Mn\textsubscript{0.2}Co\textsubscript{2}O\textsubscript{4}.

\begin{figure}[t]
\includegraphics[trim=3.7cm 10cm 5.2cm 8.1cm, clip=true,scale=0.6]{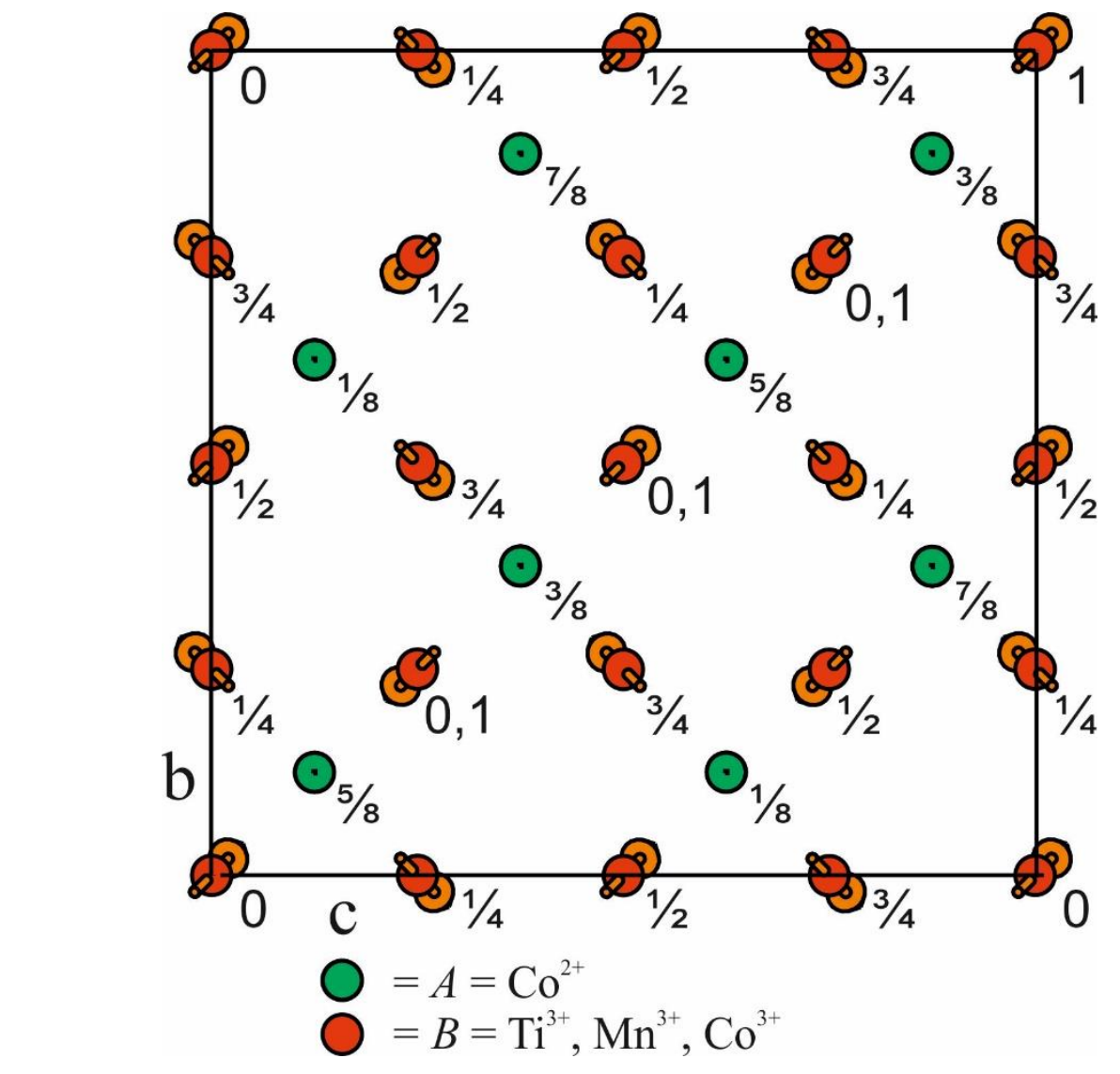}
\caption{Magnetic structure of Ti\textsubscript{0.8}Mn\textsubscript{0.2}Co\textsubscript{2}O\textsubscript{4} and Ti\textsubscript{0.6}Mn\textsubscript{0.4}Co\textsubscript{2}O\textsubscript{4}. Here we used the setting, where the moments of ferrimagnetic coupled $A$- and $B$-site atoms are aligned parallel to the a axis. Thus the noncollinear antiferromagnetic spin alignments occur in the $bc$ plane.}
\label{fig:Fig4}
\end{figure}

\begin{figure}[b]
\includegraphics[trim=0.5cm 0.5cm 5.3cm 0.8cm, clip=true,scale=1.3]{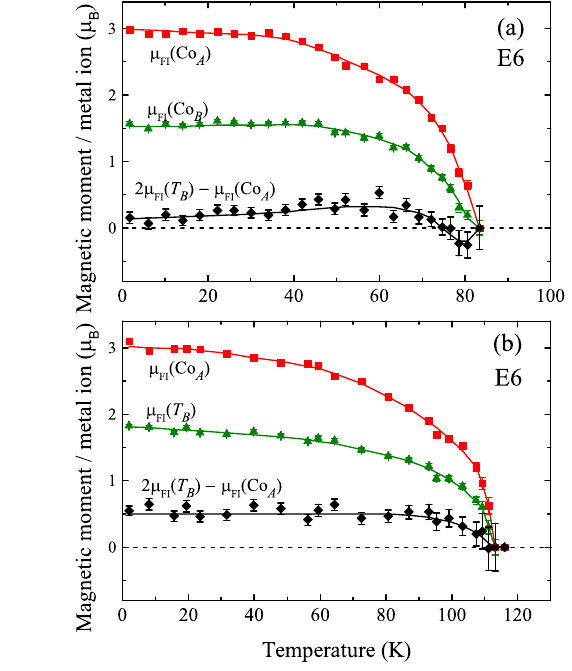}
\caption{Temperature dependence of the ferrimagnetic moments of the ions in (a) Ti\textsubscript{0.8}Mn\textsubscript{0.2}Co\textsubscript{2}O\textsubscript{4}, and (b) Ti\textsubscript{0.6}Mn\textsubscript{0.4}Co\textsubscript{2}O\textsubscript{4} located at the $A$ and $B$ sites. The $A$ site is only occupied with Co\textsuperscript{2+} ions (labelled as Co\emph{\textsubscript{A}}), while the B site contains Ti\textsuperscript{3+}, Mn\textsuperscript{3+}, and Co\textsuperscript{3+} (labelled as T\emph{\textsubscript{B}}). For the $B$ site only the averaged moments are given. The bold lines are a guide for the eye.}
\label{fig:Fig5}
\end{figure}

For Ti\textsubscript{0.6}Mn\textsubscript{0.4}Co\textsubscript{2}O\textsubscript{4} the values, which vary between 0.60 and 0.65, are found to be smaller. This may give the trend that the moment of Co\textsuperscript{3+} ion decreases with increasing Mn content. The Co\textsuperscript{2+} ion
at the \emph{A} site have three unpaired electrons in the 3\emph{d}\textsuperscript{7} configuration. Here the magnitude of magnetic moment reaches between 2.60 and 3.10 $\mu$\textsubscript{B}, which
are close to the theoretical value of the high-spin state $\mu$\textsubscript{eff} = \emph{g}\emph{S}$\mu$\textsubscript{B} = 3.0 $\mu$\textsubscript{B}. In contrast the moment of Co\textsuperscript{2+} ion at the \emph{A} site in TiCo\textsubscript{2}O\textsubscript{4} was found to be reduced which can be ascribed to stronger frustration effects \cite{Thota2017}.

In table 2 we also compare the resulting ferromagnetic components of Ti\textsubscript{0.8}Mn\textsubscript{0.2}Co\textsubscript{2}O\textsubscript{4} and Ti\textsubscript{0.6}Mn\textsubscript{0.4}Co\textsubscript{2}O\textsubscript{4} determined from magnetization measurement with those obtained from our neutron diffraction study given as 2$\mu$\textsubscript{FI}(\emph{T\textsubscript{B}}) $-$ $\mu$\textsubscript{FI}(Co\emph{\textsubscript{A}}). Due to the fact that the resulting ferromagnetic components are relatively small we were not able to determine them with good accuracy from our neutron diffraction data. Taking this into account the moment values obtained from both methods show more or less a good agreement (table 2). We further have investigated the thermal variation of the magnetic moments of Ti\textsubscript{0.8}Mn\textsubscript{0.2}Co\textsubscript{2}O\textsubscript{4}
and Ti\textsubscript{0.6}Mn\textsubscript{0.4}Co\textsubscript{2}O\textsubscript{4}. Due to the fact that the AF component could not be precisely determined from the E6 data in the temperature range close to the Curie temperature we used the ratio $\mu$\textsubscript{FI}(\emph{T\textsubscript{B}})/$\mu$\textsubscript{AF}(\emph{T\textsubscript{B}}) as constraint. As given above we were not able to determine precisely the individual moment values of Ti\textsuperscript{3+}, Mn\textsuperscript{3+}, and Co\textsuperscript{3+} at the \emph{B} site. Therefore, only the averaged moment values at the \emph{B} site (labelled as \emph{T\textsubscript{B}}) are plotted in figure 5 which are coupled ferrimagnetically with the moment values at the \emph{A} site (labelled as Co\emph{\textsubscript{A}}).

\begin{figure}[b]
\includegraphics[trim=0.5cm 0.5cm 5.3cm 0.5cm, clip=true,scale=1.2]{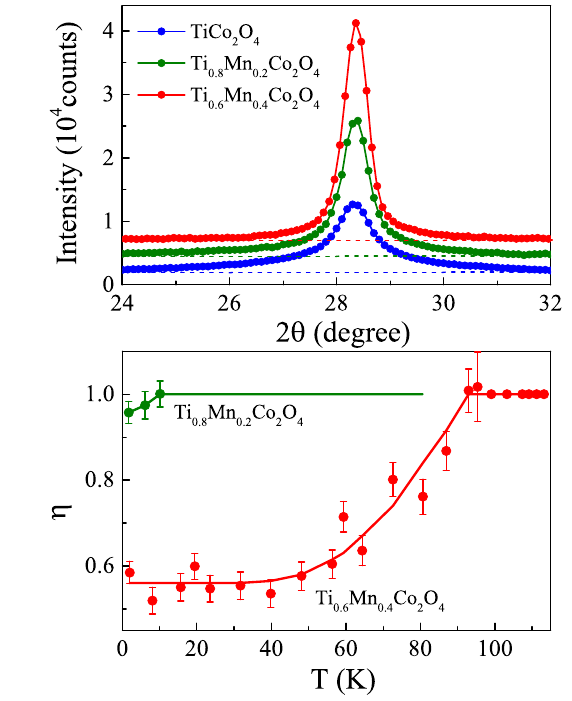}
\caption{Change of the peak shape of the magnetic reflection (101)\textsubscript{M} with increasing Mn content. A pure Lorentzian peak shape has been observed for TiCo\textsubscript{2}O\textsubscript{4}, where one also finds a broad diffuse signal giving an uneven background. The data of TiCo\textsubscript{2}O\textsubscript{4} were taken from our earlier study \cite{Thota2017}. With increasing Mn content a reduction of diffuse scattering is observed. The strong increase of the magnetic intensity of the reflection (101)\textsubscript{M} can be ascribed to the fact that the magnetic moment of Mn is much stronger than that of Ti. An increase of the Mn content also leads to change of the peak shape, where the Lorentzian part $\eta$ of the pseudo-Voigt function is decreasing. The pseudo-Voigt function is defined as \emph{pV}(\emph{x}) = $\eta$\emph{L}(\emph{x}) + (1$-$$\eta$)\emph{G}(\emph{x}). Interestingly the Lorentzian part $\eta$ is also increasing as a function of temperature shown in the lower part of the figure. From the instrumental resolution curve one expects at the position of the reflection (101)\textsubscript{M} a FWHM of 0.52\textsuperscript{o} which practically has a Gaussian shape. The bold line is a guide for the eye.}
\label{fig:Fig6}
\end{figure}

We also studied the temperature variation of the peak shape. For the Rietveld refinements we used
the pseudo-Voigt function which is defined as \emph{pV}(\emph{x}) = $\eta$\emph{L}(\emph{x}) + (1 $-$ $\eta$)\emph{G}(\emph{x}), where $\eta$ gives the contribution of the Lorentzian part. In figure 6 it can be seen that the Lorentzian part for Ti\textsubscript{0.6}Mn\textsubscript{0.4}Co\textsubscript{2}O\textsubscript{4} is strongly increasing with the temperature reaching finally a pure Lorentzian peak shape at about 90 K somewhat below the magnetic ordering temperature \emph{T}\textsubscript{C} = 112 K. In contrast, for Ti\textsubscript{0.8}Mn\textsubscript{0.2}Co\textsubscript{2}O\textsubscript{4}
the pure Lorentzian peak shape is already reached at 10 K below the compensation point. As reported earlier a pure Lorentzian peak shape has been observed for TiCo\textsubscript{2}O\textsubscript{4} in the full temperature range which was found to be broadened \cite{Thota2017}. The change
of the peak shape can be ascribed to the presence of a superimposed contribution of diffuse scattering. But it has to be mentioned that a Lorentzian peak shape may also describe the distribution of the magnetic mosaic blocks or the presence of short-range order frozen in a spin-glass state. Further studies are needed to give a detailed quantitative description of this behaviour. According to the chemical percolation approach reported by Cowley \emph{et al}. one should expect Bragg scattering occurs from the largely linked clusters, together with a diffuse scattering occurring from the limited
ferrimagnetic clusters below a critical composition \cite{mizoguchi1977measurement}. In such
case the correlation length should consist of two components, (i) transverse component which diverges across \emph{T}\textsubscript{N} and decreases below and it is associated with critical fluctuations within the infinite cluster. The second one is the (ii) longitudinal component which dominates at low temperature and it is linked to the finite clusters. It is well known that diffuse scattering is an important tool for analyzing such longitudinal and transverse spin components, in particular, the short-range-order of reentrant spin-glass systems in which the disordered reenters at low-temperature below \emph{T}\textsubscript{N}. 

In the present case a slight decrease of the A-site magnetic moment indicates the coexistence of longitudinal and transverse spin components associated with a long-range ordered magnetic network and a glassy state, respectively \cite{chakravarthy1991perturbed}. Such disordered spins do not give any contribution to magnetic Bragg intensities. Neutron diffraction studies of systems like FeMnTiO\textsubscript{3} and Fe\textsubscript{0.5}Mg\textsubscript{0.45}Cl\textsubscript{2} reported the existence of magnetic diffuse scattering due to the coexistence of antiferromagnetic long-range ordering and a transverse spin-glass state \cite{yoshizawa1987mixed,wong1985coexistence}. The appearance of superimposed magnetic diffuse scattering at the antiferromagnetic Bragg peak indicates the changes taking place in local magnetic ordering. It is also reported that the intensity of such diffuse scattering changes drastically with decreasing temperature, while the intensity of the antiferromagnetic Bragg peak slightly decreases \cite{yoshizawa1987mixed,wong1985coexistence}. In the 1980s the existence of a mixed phase (spin-glass and ferri/ferro/antiferromagnetic orderings) system lead to controversial discussion. There is a fair amount of debate on the nature of this magnetic order till date despite many advances in the condensed matter theory which predicts various forms of reentrant behavior (in both Ising and Heisenberg systems) \cite{kleemann2010coexistence,frantz2019glassy,kmjec201957fe,ghara2018coexistence}. Wong \emph{et al}. reported the coexistence of spin-glass and antiferromagnetic orders in the Ising system Fe\textsubscript{0.55}Mg\textsubscript{0.45}Cl\textsubscript{2} \cite{wong1985coexistence}. In order to probe precisely the transverse spin-glass nature of this system these authors recorded diffuse scattering scans at 1.5 K along the three different crystallographic directions below the freezing temperature ($\sim$3 K). Consequently, they observed a superimposed Lorentzian diffuse magnetic peak at the position of the nuclear Bragg peak with the correlation length $\xi$ $\sim$ 10 \AA. Such frozen short-range correlation is probably the source of the spin-glass behavior observed in single crystals of Fe\textsubscript{0.55}Mg\textsubscript{0.45}Cl\textsubscript{2} \cite{wong1985coexistence}. Considering these results, in the present case we performed a detailed ac-magnetic susceptibility and diffuse neutron scattering studies to probe the reentrant glassy behavior.

Figure 7 shows an enlarged view around the base of 111 neutron-diffraction peak of Ti\textsubscript{0.8}Mn\textsubscript{0.2}Co\textsubscript{2}O\textsubscript{4} and Ti\textsubscript{0.6}Mn\textsubscript{0.4}Co\textsubscript{2}O\textsubscript{4}, well below (at 1.7 K/2 K) and above (at 90 K/142 K) the long-range ordering temperature revealing the presence of diffuse scattering superimposed on the reflection 200. Such well-resolved Bragg reflections, originating from the spatial ordering of the transverse-spin components (TSC), indicates the presence of canted local spin configuration on the \emph{B} sublattice (similar to the Yafet-Kittel's (Y-K) three sublattice model)\cite{hubsch1982semi,murthy1969yafet,chakravarthy1991perturbed}. 

\begin{figure}[b]
\includegraphics[trim=-0.7cm 0.5cm 4.8cm 0.2cm, clip=true,scale=1.1]{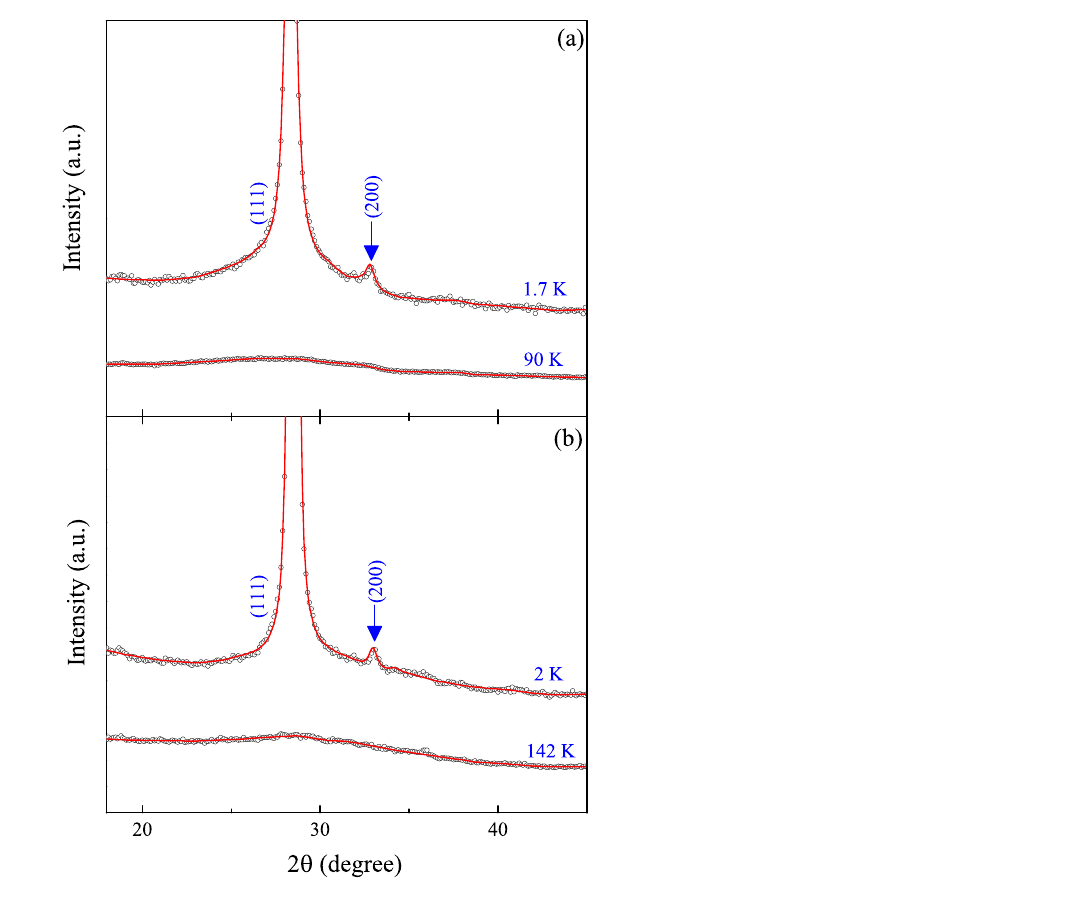}
\caption{Temperature dependence zoomed view of neutron diffraction pattern around the 200 peak for (a) \emph{x} = 0.2 and (b) \emph{x} = 0.4.}
\label{fig:Fig7}
\end{figure}

Usually, in a non-collinear system, the individual site moments deduced from the neutron diffraction study are due to the longitudinal spin components of the canted site moments, whereas the transverse components do not contribute to the normal Bragg reflections. In a triangular type spin-lattice Bragg reflections arise due to the spatial ordering of the transverse spin components. Thus, the origin of the Bragg reflection 200 indicates the presence of a canted local spin configuration on the \emph{B} sublattice. These results are consistent with the previous reports \cite{murthy1969yafet,chakravarthy1991perturbed}. We have evaluated the canting angle ($\alpha$\textsubscript{Y-K}) using the individual \emph{B}-site moments (AF and FI) derived from the neutron diffraction study of Ti$_{1-x}$Mn$_{x}$Co\textsubscript{2}O\textsubscript{4} with \emph{x} = 0.2 and \emph{x} = 0.4. Accordingly, the canting angle is $\alpha$\textsubscript{Y-K} lies between 66.9 and 73.7$\degree$ for \emph{x} = 0.2, and 71 and 74.1$\degree$ for \emph{x} = 0.4. Such non-collinear behavior of spins causes substantial decrease of the magnitude of \emph{B}-sublattice moments than the expected value and frequency dispersion of the dynamic susceptibility (which will be discussed in
later sections). Usually, the transverse spin-glass component (often called `semi-spin-glass') coexists with the longitudinal-spin component when the magnetic ions are diluted with nonmagnetic ions like
Ti\textsuperscript{4+}, Sn\textsuperscript{4+} or Zn\textsuperscript{2+}. Because of such dilution and magnetic frustration the spins in the \emph{B} sublattice may cant locally \cite{villain1979insulating}. Theoretical studies reported by Villain show that two distinct transitions may occurs in such semi-spin glass systems. The first one is the Néel temperature \emph{T}\textsubscript{N} corresponding to the breakdown of LSC and the second transition is the spin-glass freezing temperature \emph{T}\textsubscript{F} at which the TSC freezes-in \cite{villain1979insulating}.

\subsection{\label{sec:level3}Electronic structure:}
The electronic and chemical state of the elements in Ti$_{1-x}$Mn$_{x}$Co\textsubscript{2}O\textsubscript{4} polycrystalline samples was analyzed by means of X-ray photoelectron spectroscopy (XPS). Figure 8 shows the detailed XPS spectra of Ti\textsubscript{0.6}Mn\textsubscript{0.4}Co\textsubscript{2}O\textsubscript{4} sample for (\emph{i}) Mn-2\emph{p}, (\emph{ii}) Co-2\emph{p},
(\emph{iii}) Ti-2\emph{p} and (\emph{iv}) O-1\emph{s} core level photoelectrons whose photoelectron intensity was plotted as a function of binding energy (eV). All these spectra were calibrated by selecting the binding energy of carbon C-1\emph{s} orbital (located at \emph{E}\textsubscript{C} = 284.8 eV) as an internal reference. The Mn-2\emph{p} spectrum of octahedrally coordinated `Mn' ion in
Ti\textsubscript{0.6}Mn\textsubscript{0.4}Co\textsubscript{2}O\textsubscript{4} spinel is shown in figure 8(a) which requires a minimum of three peaks (two main peaks at 640.45 eV and 652.12 eV, and one broad satellite peak at 628.52 eV) to reproduce the spectrum. The peak component of Mn is constrained to have the same peak profile within the binding energy range of 0.2 eV. The binding-energy separation between the Mn doublet $\delta$\emph{E}\textsubscript{Mn} (\emph{E\textsubscript{p}}\textsubscript{1/2} $-$ \emph{E\textsubscript{p}}\textsubscript{3/2}) $\sim$ 11.67 eV signifies the trivalent oxidation state of Mn inside the core of spinel structure Ti$_{1-x}$Mn$_{x}$Co\textsubscript{2}O\textsubscript{4} \cite{kim2012electronic,pramanik2017effects,wei2016tuning}.
 
\begin{figure}[t]
\includegraphics[trim=0.2cm 0cm 0cm 0cm, clip=true,scale=0.8]{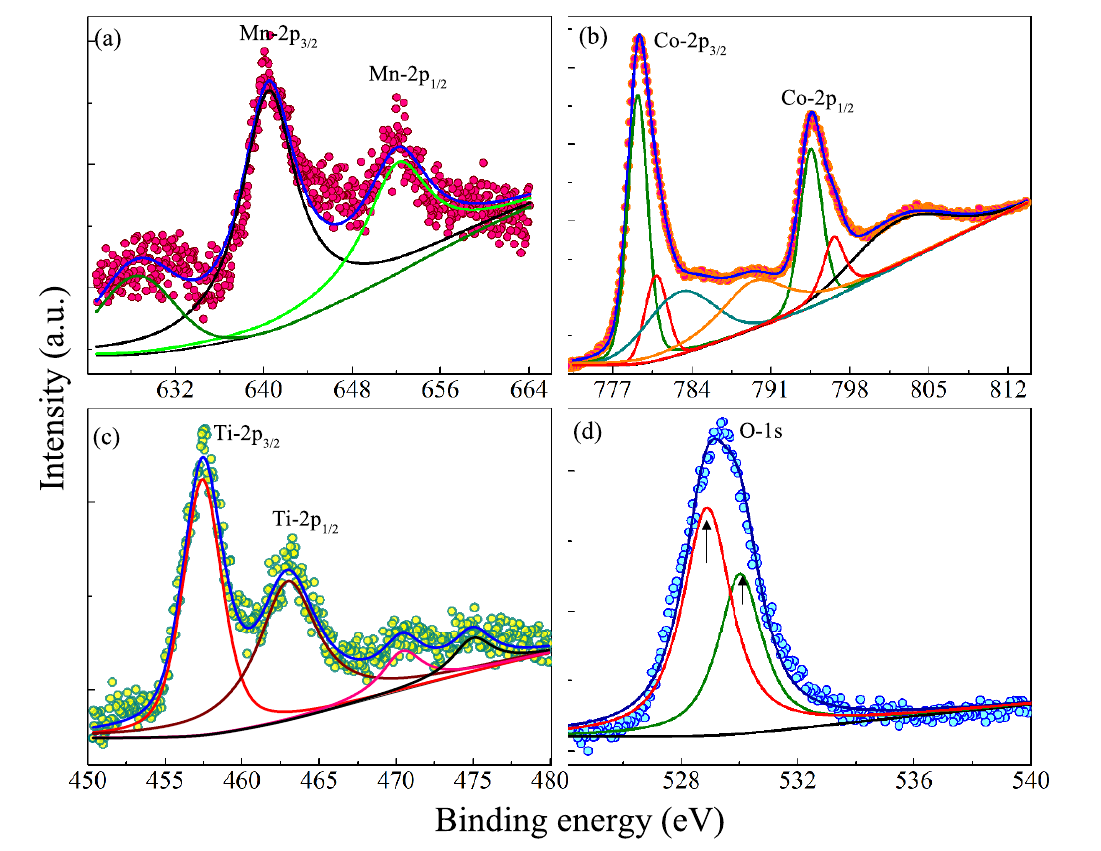}
\caption{Figure 8. The X-ray photoelectron spectra of Ti\textsubscript{0.6}Mn\textsubscript{0.4}Co\textsubscript{2}O\textsubscript{4}. (a), (b), (c) and (d) show Mn-2\emph{p}, Co-2\emph{p}, Ti-2\emph{p} and O-1\emph{s} core level XPS spectra respectively.}
\label{fig:Fig8}
\end{figure}

The observed value of $\delta$\emph{E}\textsubscript{Mn} in the present investigating system is slightly higher ($\sim$ 0.46 eV) as compared to undoped MnCo\textsubscript{2}O\textsubscript{4}
\cite{pramanik2017effects}. Such higher value of $\delta$\emph{E}\textsubscript{Mn} is associated
with the increase in the screening strength of Mn\textsuperscript{3+} ion by the additional Ti\textsuperscript{3+} ion present in the Ti$_{1-x}$Mn$_{x}$Co\textsubscript{2}O\textsubscript{4} matrix.

The core level XPS spectrum of Co-2\emph{p} consist of two doublets of Co(II) (at 794.6 eV) and Co(III) (779.3 eV) together with three broad satellite peaks (as shown in figure 8(b)) positioned at 783.4, 789.9 and 804.1 eV. For the peak fitting, we have imposed a constraint that the ratio between the areas of the Co-2\emph{p}\textsubscript{3/2} and Co-2\emph{p}\textsubscript{1/2} for Co(II) and Co(III) should be same. This constrain is relaxed at the final step of the peak fitting. From the fitting analysis, we have calculated the spin-orbit splitting energy ($\delta$\emph{E}) between the doublets
$\delta$\emph{E}\textsubscript{Co3+}(\emph{2p}\textsubscript{1/2} $-$ \emph{2p}\textsubscript{3/2}) and
$\delta$\emph{E}\textsubscript{Co2+}(\emph{2p}\textsubscript{1/2} $-$ \emph{2p}\textsubscript{3/2}), which comes out to be 15.3 and 15.78 eV, respectively, signifying the two different oxidation states of Co (trivalent and divalent states) \cite{Thota2017,kim2000analysis}.

On the other hand, Ti-2\emph{p} core level spectrum (figure 8(c)) is deconvoluted into two major peaks (doublet) located at 457.44 and 463.01 eV, together with two high-energy shake up satellite peaks centered at 470.46 and 475.04 eV. The initial fitting parameters are obtained from the NIST XPS Database \cite{NIST}. The intensity ratio of the Ti-2\emph{p}\textsubscript{3/2} and Ti-2\emph{p}\textsubscript{1/2} peaks is also constrained to 2:1 ratio \cite{NIST}. The Ti-2\emph{p}\textsubscript{1/2} peak profile is slightly broader ($\sim$0.28 eV) than the Ti-2\emph{p}\textsubscript{3/2} peak due to the Coster-Kronig effect \cite{biesinger2010resolving}. The magnitude of spin-orbit splitting $\delta$\emph{E}\textsubscript{Ti} between the Ti-2\emph{p}\textsubscript{1/2} and Ti-2\emph{p}\textsubscript{3/2} ($\sim$ 5.58 eV) signifies the presence of trivalent electronic state of Ti inside Ti\textsubscript{0.6}Mn\textsubscript{0.4}Co\textsubscript{2}O\textsubscript{4}
\cite{biesinger2010resolving}. In order to confirm the Ti\textsuperscript{3+} oxidation state, we have carried out a systematic correlation between the Ti-O bond length (obtained from the Rietveld refinement data) and a robust parameter $\delta$\textsubscript{O-Ti} \emph{i.e.} the binding energy
difference between the O-1\emph{s} and Ti-2\emph{p}\textsubscript{3/2} \cite{atuchin2006ti}. The experimentally observed value of $\delta$\textsubscript{O-Ti} $\sim$ 72.6 and the Ti-O bond length $\sim$ 2.058 \AA confirm the presence of Ti\textsuperscript{3+} oxidation state. Had Ti existed in tetravalent oxidation sate, the values of $\delta$\textsubscript{O-Ti} and Ti-O bond length would lie in the range 71 eV $\leq$ $\delta$\textsubscript{O-Ti} $\leq$ 72 eV and 1.94 \AA $\leq$ Ti-O $\leq$ 1.97 \AA, respectively \cite{atuchin2006ti}. For divalent oxidation state of Ti the aforementioned
parameters would be $\delta$\textsubscript{O-Ti} $\sim$ 75 eV with Ti-O bond length $\sim$2.08 \AA.

Finally, the O-1\emph{s} spectrum is deconvoluted into two partially resolved Gaussian-Lorentzian peaks centered at 530.03 and 528.88 eV as shown by arrow marks in figure 8(d). The main origin of highest
intensity peak at 528.88 eV is associated with the bonding between metal and lattice oxygen (Mn-O, Ti-O and Co-O), while the low intensity peak at 530.03 eV is associated with the surface-adsorbed oxygen \cite{kim2000analysis}. The asymmetric behavior observed in O-1\emph{s} core level spectrum is
mainly associated with the presence of oxygen vacancies and different atomic environment faced by the O$^{2-}$ anions \cite{aswaghosh2016defect}.

\subsection{\label{sec:level4}Thermal variation of Magnetic-Susceptibilities and Specific-Heat:}
In this section we present the temperature variation of dc-magnetic susceptibilities $\chi$(\emph{T}) of both Ti\textsubscript{0.6}Mn\textsubscript{0.4}Co\textsubscript{2}O\textsubscript{4} (figure 9) and
Ti\textsubscript{0.8}Mn\textsubscript{0.2}Co\textsubscript{2}O\textsubscript{4} (figure 10) measured under zero-field-cooled (ZFC) and field-cooled (FC) conditions. We have performed these measurements in the presence of different external dc-magnetic fields \emph{H}\textsubscript{dc} between 500 Oe and 50 kOe, however, figures 9 and 10 show the variation of $\chi$(\emph{T}) at \emph{H}\textsubscript{dc} = 500 Oe (figures 9(a) and 10(a)) and 50 kOe (figures 9(b) and 10(b)) for Ti\textsubscript{0.6}Mn\textsubscript{0.4}Co\textsubscript{2}O\textsubscript{4} and Ti\textsubscript{0.8}Mn\textsubscript{0.2}Co\textsubscript{2}O\textsubscript{4}, respectively. For Ti\textsubscript{0.6}Mn\textsubscript{0.4}Co\textsubscript{2}O\textsubscript{4} the $\chi$(\emph{T}) curves show ferrimagnetic behavior with a giant bifurcation ($\Delta$$\chi$ $\sim$ 7.74$\times$10\textsuperscript{-3} emu/gOe) between the $\chi$\textsubscript{ZFC}(\emph{T}) and $\chi$\textsubscript{FC}(\emph{T}) due to very high magnetocrystalline anisotropy. The ferrimagnetic ordering is suggested to result from unequal/opposite magnetic moments of cations occupied at the tetrahedral \emph{A} site {[}$\mu_{\downarrow}$(Co\textsuperscript{2+}){]} and the octahedral \emph{B} site {[}$\mu_{\uparrow}$(Mn\textsuperscript{3+},
Ti\textsuperscript{3+} and Co\textsuperscript{3+}){]}.

\begin{figure}[t]
\includegraphics[trim=0.5cm 0.2cm 0cm 0.8cm, clip=true,scale=0.77]{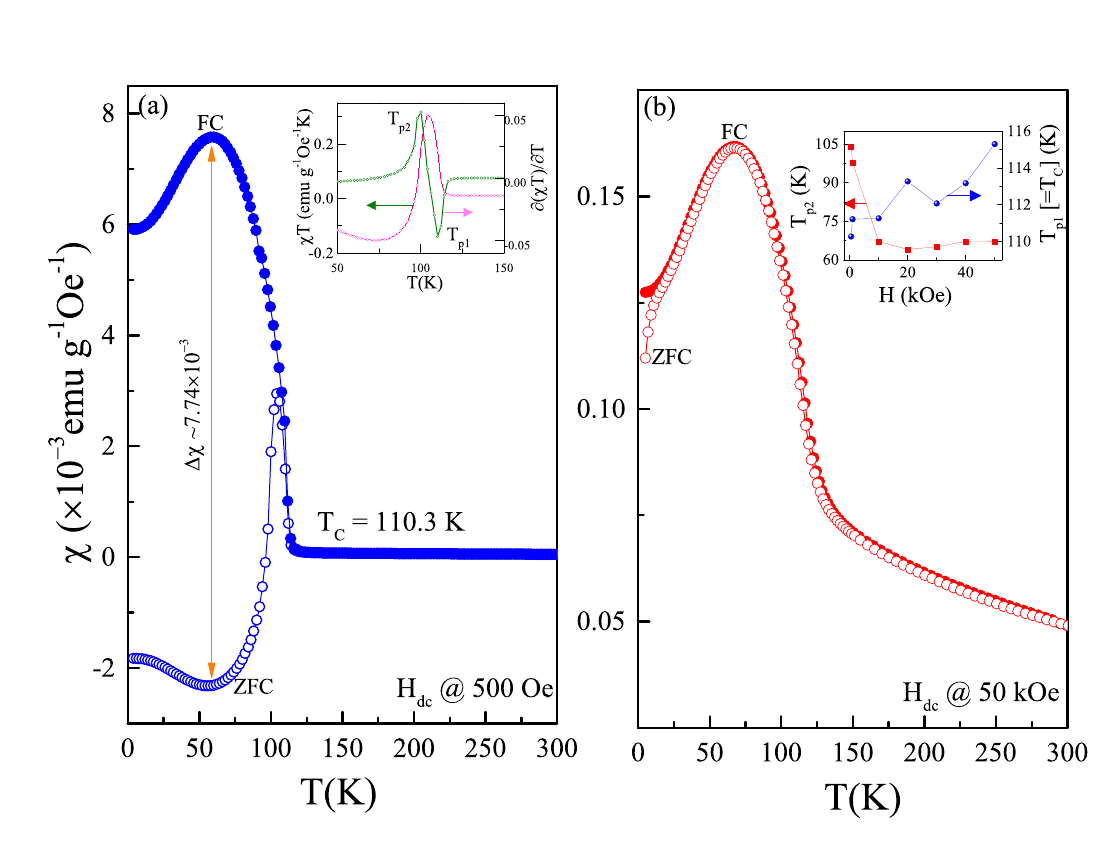}
\caption{Temperature dependence of dc-magnetic susceptibility $\chi$(\emph{T}) of Ti\textsubscript{0.6}Mn\textsubscript{0.4}Co\textsubscript{2}O\textsubscript{4} measured under zero-field-cooled (ZFC) and field-cooled (FC) conditions in the presence of external dc-magnetic field \emph{H}\textsubscript{dc} is equivalent to (a) 500 Oe and (b) 50 kOe. The inset of (a) depict product $\chi$\emph{T} (left hand side scale) and its derivative (right hand side scale) was plotted as a function of temperature.}
\label{fig:Fig9}
\end{figure}

\begin{figure}[t]
\includegraphics[trim=0.3cm 0.2cm 0cm 0.5cm, clip=true,scale=0.76]{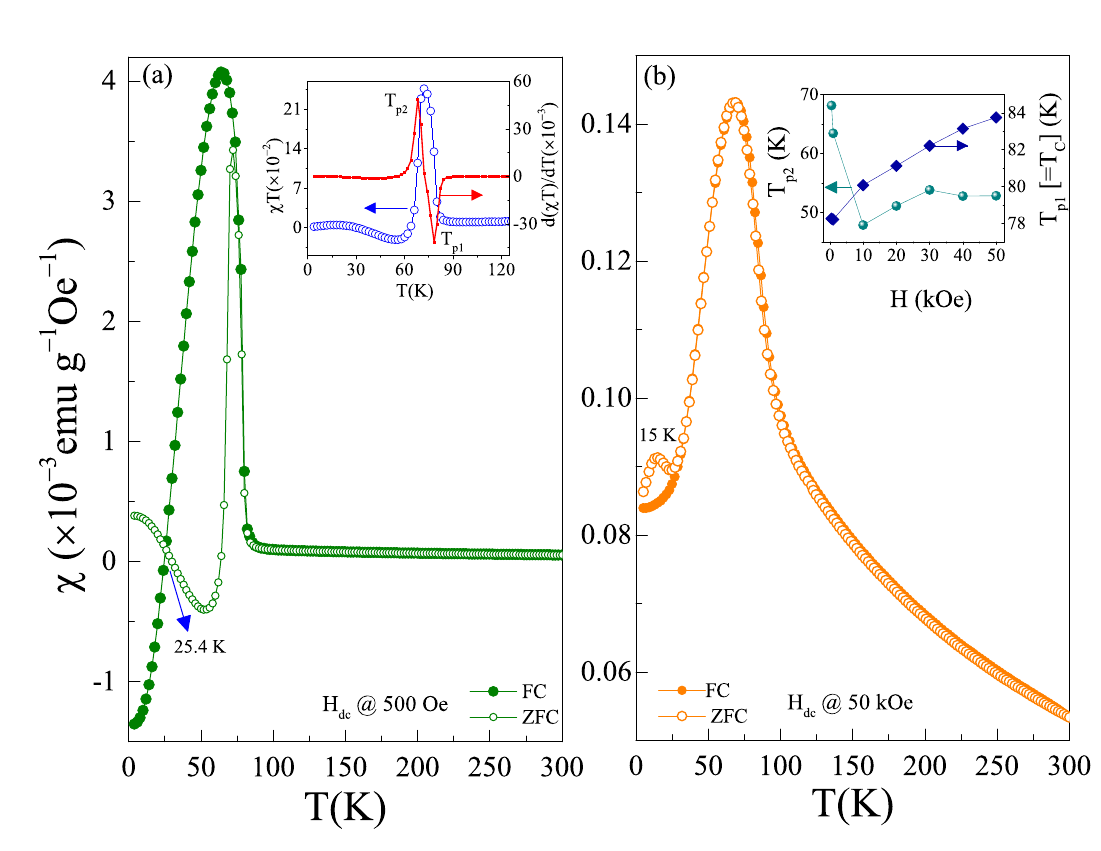}
\caption{Temperature dependence of dc-magnetic susceptibility $\chi$(\emph{T}) of Ti\textsubscript{0.8}Mn\textsubscript{0.2}Co\textsubscript{2}O\textsubscript{4} measured under zero-field-cooled (ZFC) and field-cooled (FC) conditions in the presence of external dc-magnetic field \emph{H}\textsubscript{dc} is equivalent to (a) 500 Oe and (\emph{b}) 50 kOe. Figure (\emph{a}) clearly shows the magnetic compensation phenomena at \emph{T}\textsubscript{COMP} = 25.4 K, represented by blue arrow mark in the main panel. However, such \emph{T}\textsubscript{COMP} disappears at very high fields figure (\emph{b}). The inset of (\emph{a}) depict product $\chi$\emph{T} (left hand side scale) and its derivative (right hand side scale) was plotted as a function of temperature.}
\label{fig:Fig10}
\end{figure}

For the composition, Ti\textsubscript{0.6}Mn\textsubscript{0.4}Co\textsubscript{2}O\textsubscript{4}
both $\chi$\textsubscript{ZFC}(\emph{T}) and $\chi$\textsubscript{FC}(\emph{T}) exhibit peaks at 104.1 K (\emph{T}\textsubscript{p-zfc}) and 60 K (\emph{T}\textsubscript{p-fc}), and drops to zero across the
\emph{T}\textsubscript{C} $\sim$ 110.3 K. It is interesting to note that $\chi$\textsubscript{ZFC}(\emph{T}) exhibits negative values for all the temperatures below 96 K with a negative maximum
($-$2.32$\times$10\textsuperscript{-3} emu/gOe) at 56.1 K, while the magnitude of $\chi$\textsubscript{FC} decreases significantly below \emph{T}\textsubscript{p-fc} without any negative values. Such markedly different characteristics of $\chi$\textsubscript{FC}(\emph{T}) and $\chi$\textsubscript{ZFC}(\emph{T}) below their peak values (at \emph{T}\textsubscript{p-zfc} and \emph{T}\textsubscript{p-fc}) are mainly due to the different temperature dependence of the magnetic
moments associated with the cations occupying \emph{A} and \emph{B} sites. For the accurate determination of the ordering temperature we have determined the susceptibility derivatives (based on the Fisher's relation which links the heat capacity and susceptibility derivative \cite {fisher1962relation}). Accordingly, we have plotted the temperature dependence of the product $\chi$\emph{T} (left hand side scale) and its derivative d($\chi$\emph{T})/d\emph{T} (right hand side scale) in the insets of figures 9(a) and 10(a). Interestingly we noticed two peaks \emph{T}\textsubscript{P1} and \emph{T}\textsubscript{P2} in the d($\chi$\emph{T})/d\emph{T} plotted as a function of temperature. The peak position \emph{T}\textsubscript{P1} is close to 110.27 K for 500 Oe signifying the exact value of \emph{T}\textsubscript{C} (consistent with hump noticed at 109.1 K from \emph{C\textsubscript{P}} vs. \emph{T} and \emph{C}\textsubscript{P}\emph{T}\textsuperscript{-1} plots shown in figure 11), whereas for the high fields (50 kOe) much broader \emph{T}\textsubscript{P1} (115.3 K) and \emph{T}\textsubscript{P2} (50.7 K) are noticed.

On the other hand, a remarkable change in the $\chi$(\emph{T}) behavior was observed with decreasing Ti\textsuperscript{3+}-ion concentration at the \emph{B} site of the spinel lattice. Magnetic compensation effect at 25.4 K (\emph{T}\textsubscript{COMP}), negative magnetization in \emph{M}\textsubscript{FC} below \emph{T}\textsubscript{COMP} and drastic decrease in \emph{T}\textsubscript{C} (82 K) from ($\chi$ $-$ \emph{T}) plot for \emph{H}\textsubscript{dc} = 500 Oe) are the noteworthy
features observed for \emph{x} = 0.2 system (figure 10). It is well known that magnetic compensation occurs in systems where the two sublattices magnetization balances with each other (similar to
Co\textsubscript{2}TiO\textsubscript{4}, \emph{T}\textsubscript{COMP} = 30.4 K) \cite{Thota2017}. Below the compensation point (\emph{T}\textsubscript{COMP}) an opposite trend was observed between $\chi$\textsubscript{FC} ($-$ve values) and $\chi$\textsubscript{ZFC} (+ve values). For \emph{x} = 0.2, both $\chi$\textsubscript{FC}(\emph{T}) and $\chi$\textsubscript{ZFC}(\emph{T}) exhibits cusp like behavior with the peak positions at 72.1 (\emph{T}\textsubscript{p-zfc}) and 64 K
(\emph{T}\textsubscript{p-fc}) corresponding to the maximum susceptibility values
$\chi$\textsubscript{Max}(\emph{T}\textsubscript{p-zfc}) = 3.42$\times$10\textsuperscript{$-$3} emu/gOe and $\chi$\textsubscript{Max}(\emph{T}\textsubscript{p-fc}) = 4.07$\times$10\textsuperscript{$-$3} emu/gOe, respectively. Without any Mn contribution (i.e. for \emph{x} = 0 case Co\textsubscript{2}TiO\textsubscript{4} system) the magnetic compensation occurs at 30.4 K with quasi long-range ferrimagnetic ordering across 48.6 K. 

Nevertheless, the moments $\mu$\textsubscript{FI}(Co\emph{\textsubscript{A}}) and $\mu$(\emph{T\textsubscript{B}}) obtained from the neutron experiments gradually decreases to zero in consonance with the $\chi$\textsubscript{dc}(\emph{T}), however, we could not able to see any negative moment or any anomaly in temperature dependence of magnetic moments (figure 5a) across the compensation point.  Such discrepancy is arising due to the fact that the neutron diffraction measurements were performed in zero-field, whereas the dc magnetization experiments were performed in the presence of externally applied magnetic field. Moreover, the present neutron diffraction results demonstrate the unequal change of the magnetic moments of the A- and B-site cations. Their different temperature dependence only plays a key role in realizing the magnetization reversal in the \emph{x} = 0.2 composition.

\begin{figure}[b]
\includegraphics[trim=0.5cm 0.2cm 0cm 1cm, clip=true,scale=0.79]{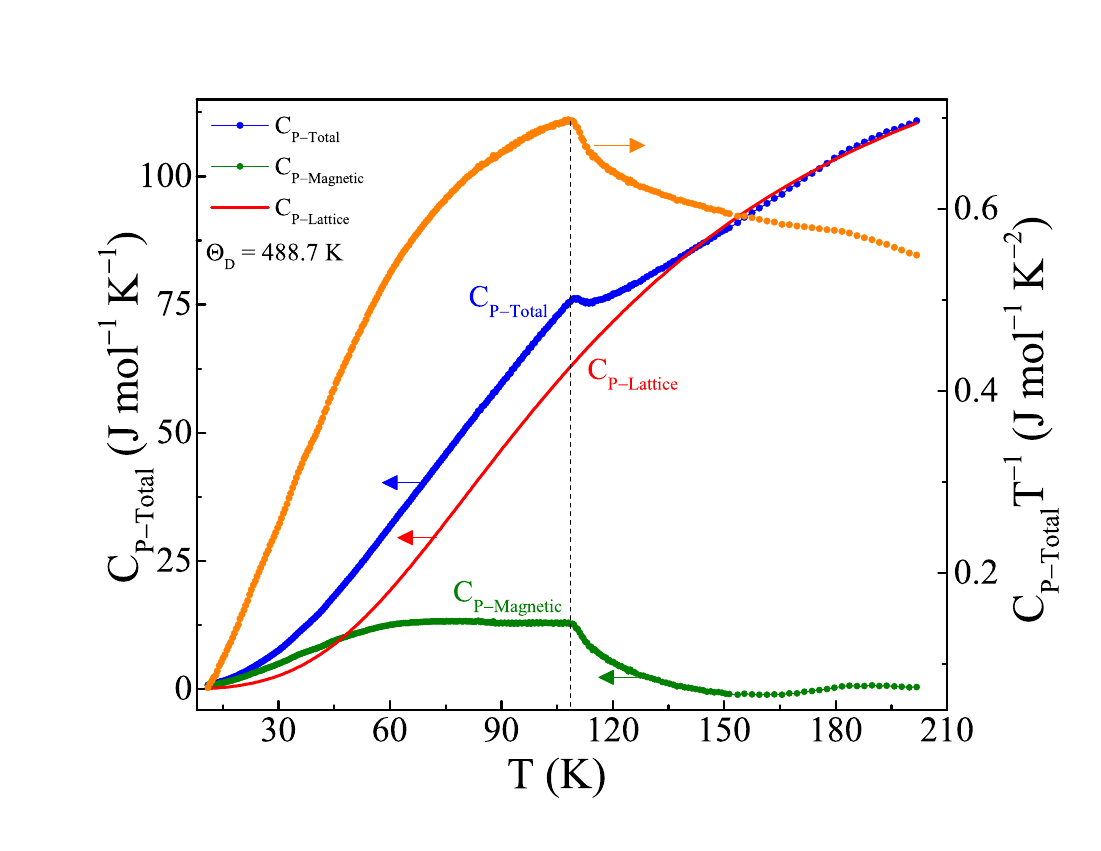}
\caption{Temperature dependent total specific-heat \emph{C\textsubscript{P}}\textsubscript{-Total}(\emph{T}) (blue color circular symbols shown on L.H.S scale) of Ti\textsubscript{0.6}Mn\textsubscript{0.4}Co\textsubscript{2}O\textsubscript{4} polycrystalline samples measured under zero field condition. The red color solid line represents the lattice contribution of specific heat calculated from numerical fits using equation \emph{C\textsubscript{P}}\textsubscript{-Phonon}=\emph{Nf}\textsubscript{D}($\Theta$\textsubscript{D}/\emph{T})=9\emph{N}\emph{R}(\emph{T}/$\Theta$\textsubscript{D})\textsuperscript{3}\(\int_{0}^{\frac{\theta_{D}}{T}}{\frac{x^{4}e^{x}}{{(e^{x} - 1)}^{2}}\text{\ dx\ }})\) as described in the text and green solid circular symbols represent the magnetic contribution to the specific heat. The R.H.S scale shows the temperature dependence of \emph{C\textsubscript{P}}/\emph{T} showing the transition at 109.1 K clearly.}
\label{fig:Fig11}
\end{figure}

For both the compounds Ti\textsubscript{0.6}Mn\textsubscript{0.4}Co\textsubscript{2}O\textsubscript{4}
and Ti\textsubscript{0.8}Mn\textsubscript{0.2}Co\textsubscript{2}O\textsubscript{4} the applied field \emph{H}\textsubscript{dc} plays a major role on the global magnetic ordering, in particular, the following are the noteworthy features (shown in figures 9(b), 10(b): (\emph{i}) disappearance of the giant bifurcation between $\chi$\textsubscript{zfc} and $\chi$\textsubscript{fc}, (\emph{ii}) unusual broadening and drastic decrease of the \emph{T}\textsubscript{p2}, (\emph{iii}) shift of \emph{T}\textsubscript{p1} (\emph{T}\textsubscript{C}) to higher temperatures with significant broadening as expected for a typical ferrimagnetic transition, and (\emph{iv}) emergence of a field induced
transition at 15 K in $\chi$\textsubscript{zfc} (but for \emph{T} \textgreater{} 73.3 K both $\chi$\textsubscript{zfc} and $\chi$\textsubscript{fc} merging with each other) and vanishing of \emph{T}\textsubscript{COMP} and negative magnetization. The inset of figures 9(b) and 10(b) depicts the variation of \emph{T}\textsubscript{p1} (= \emph{T}\textsubscript{C}) and \emph{T}\textsubscript{p2} as a function of applied field \emph{H}\textsubscript{dc} estimated from the derivative analysis {[}d($\chi$\emph{T})/d\emph{T}{]}. These plots clearly show the increasing trend of \emph{T}\textsubscript{p1} (from 110.27 K to 115.3 K for \emph{H}\textsubscript{dc} = 500 Oe to 50 kOe,
respectively) with an anomaly across 113 K (for \emph{H}\textsubscript{dc} = 20 kOe) for \emph{x} = 0.4. Whereas, \emph{T}\textsubscript{p2} decreases drastically from 104 K to 64 K for
\emph{H}\textsubscript{dc} = 0.5 and 20 kOe, respectively and increases thereafter to 67.21 K (for 50 kOe) indicating the loosely bound spins across \emph{T}\textsubscript{p2}. Such condition is quite similar to TiCo\textsubscript{2}O\textsubscript{4} in which reentrant spin-glass like characteristics was reported just below the ferrimagnetic ordering \cite{Thota2017}. Further no sharp peak observed in \emph{C}\textsubscript{P} across \emph{T}\textsubscript{C} and a clear drop in the magnitude of \emph{C\textsubscript{P}}/\emph{T} at 109.1 K suggesting the entropy loss due to disordering of spins (likely a spin-glass-like state) just below the \emph{T}\textsubscript{C}. The left-hand-side scale of figure 11 shows the temperature dependence of the total specific heat
\emph{C\textsubscript{P}}\textsubscript{-Total}(\emph{T}) and the individual contributions from lattice
{[}\emph{C\textsubscript{P}}\textsubscript{-Lattice}(\emph{T}){]} and magnetic        
{[}\emph{C\textsubscript{P}}\textsubscript{-Magnetic}(\emph{T}){]} specific heats for the polycrystalline sample of Ti\textsubscript{0.6}Mn\textsubscript{0.4}Co\textsubscript{2}O \textsubscript{4}. Usually, \emph{C\textsubscript{P}}\textsubscript{-Total}(\emph{T}) of a crystalline system is comprised of the magnetic and lattice contributions. In particular, the contribution of
\emph{C\textsubscript{P}}\textsubscript{-Lattice} is a sum of both electronic \emph{C\textsubscript{P}}\textsubscript{-electron} and phonon \emph{C\textsubscript{P}}\textsubscript{-Phonon} parts. Since the
electronic contribution is significant only at very low temperatures, the phonon involvement can be extracted from the following expression \cite{bouvier1991specific}:

\begin{equation}
\begin{split}
\emph{C\textsubscript{P}}\textsubscript{-Phonon}=\emph{Nf}\textsubscript{D}(\Theta\textsubscript{D}/\emph{T})\\
=9\emph{N}\emph{R}(\emph{T}/\Theta\textsubscript{D})\textsuperscript{3}\int_{0}^{\frac{\theta_{D}}{T}}{\frac{x^{4}e^{x}}{{(e^{x} - 1)}^{2}}dx}
\end{split}
\end{equation}

\begin{figure}[t]
\includegraphics[trim=6cm 13cm 6cm 3cm, clip=true,scale=0.8]{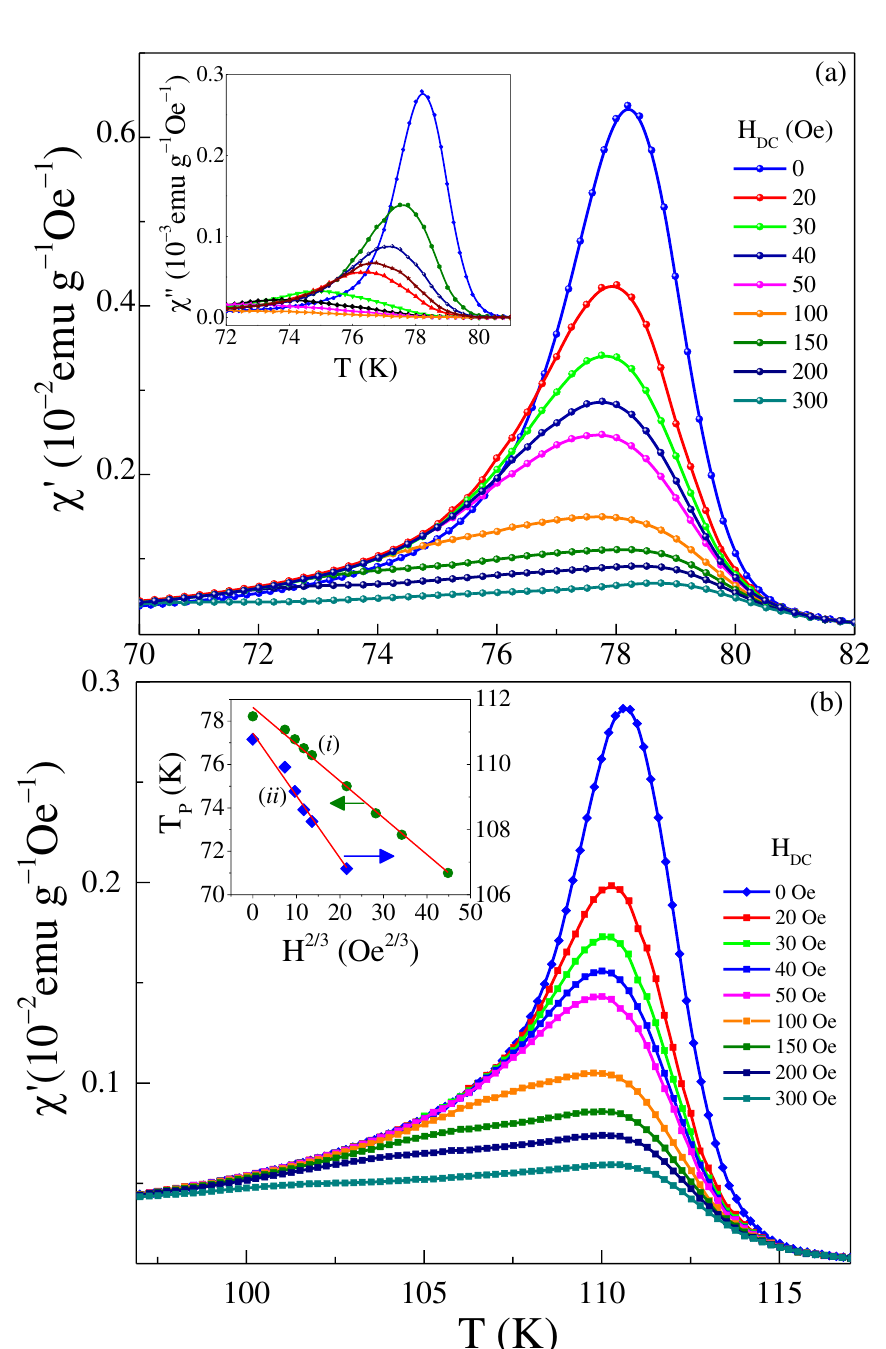}
\caption{The temperature dependence of real $\chi$\textsuperscript{$\prime$}(\emph{T}) and imaginary $\chi$\textsuperscript{$\prime\prime$}(\emph{T}) components of ac-magnetic susceptibility $\chi$\textsubscript{ac}(\emph{T}) of the compositions (a) Ti\textsubscript{0.8}Mn\textsubscript{0.2}Co\textsubscript{2}O\textsubscript{4}, and (b) Ti\textsubscript{0.6}Mn\textsubscript{0.4}Co\textsubscript{2}O\textsubscript{4} recorded at various dc magnetic fields (0 $\textless$ \emph{H}\textsubscript{dc} $\textless$ 300 Oe) with a constant ac-field of peak-to-peak amplitude \emph{h}\textsubscript{ac} $\sim$ 4 Oe and frequency $f$ $\sim$ 51 Hz. The variation of peak maximum temperature obtained from $\chi$\textsuperscript{$\prime\prime$}(\emph{T}) plotted as a function of \emph{H}\textsubscript{dc}\textsuperscript{2/3} for both the compositions (\emph{i}) Ti\textsubscript{0.8}Mn\textsubscript{0.2}Co\textsubscript{2}O\textsubscript{4}, and (\emph{ii}) Ti\textsubscript{0.6}Mn\textsubscript{0.4}Co\textsubscript{2}O\textsubscript{4} are shown in the inset figure of (b).}
\label{fig:Fig12}
\end{figure}

\begin{figure*}[t]
\includegraphics[trim=-1.4cm 2.9cm 0cm 0.4cm, clip=true,scale=1.3]{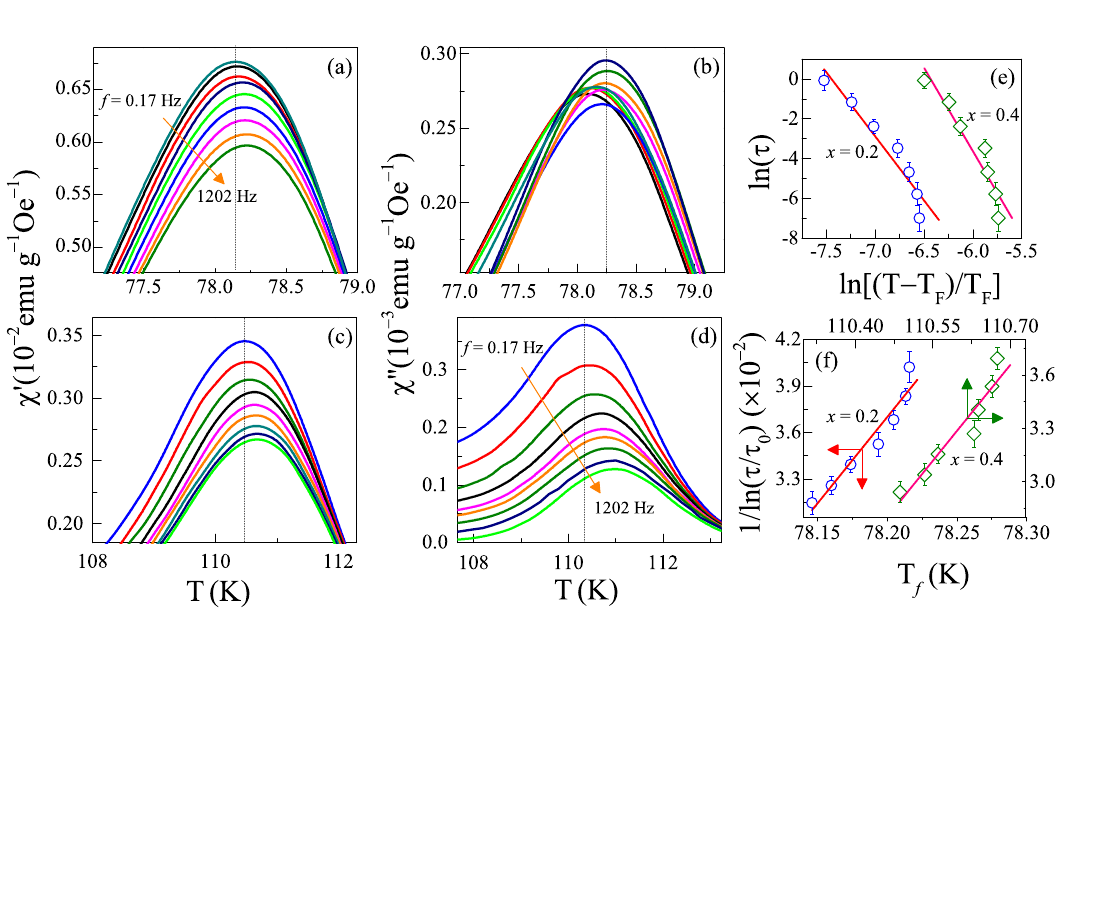}
\caption{The zoomed view of temperature dependence of real $\chi$\textsuperscript{$\prime$}(\emph{T}) and imaginary $\chi$\textsuperscript{$\prime\prime$}(\emph{T}) components of ac-magnetic-susceptibility measured at different frequencies ($f$) in the range 0.17 $\textless$ $f$ $\textless$ 1202 Hz with \emph{h}\textsubscript{ac} = 4 Oe and \emph{H}\textsubscript{dc} = 0 under warming condition for $x$ = 0.2 (a and b) and $x$ = 0.4 (c and d). ln($\tau$) versus ln[(\emph{T}$-$\emph{T}\textsubscript{F})/\emph{T}\textsubscript{F}] (e), and 1/ln($\tau$/$\tau$\textsubscript{o}) versus \emph{T}$_{f}$ (f) corresponding to the Power-law (PL) and Vogel-Fulcher law (VFL), respectively for both the compounds. The solid-lines represent the best fits of VFL and PL to the circular and diamond symbols (represent the peak temperature obtained from $\chi$\textsuperscript{$\prime$}), respectively for both the systems.}
\label{fig:Fig13}
\end{figure*}

In the above equation, the quantity \emph{f}\textsubscript{D}($\Theta$\textsubscript{D}/\emph{T}) (=9\emph{R}(\emph{T}/$\Theta$\textsubscript{D})\textsuperscript{3}\(\int_{0}^{\frac{\theta_{D}}{T}}{\frac{x^{4}e^{x}}{{(e^{x} - 1)}^{2}}\text{dx}}\)) represents Debye function, where \emph{N} is the number of atoms per formula unit, \emph{R} is the universal gas constant (8.314 J/mol K), and $\Theta$\textsubscript{D} is the Debye temperature. To evaluate the \emph{C}\textsubscript{P-Phonon}
contribution and $\Theta$\textsubscript{D,} we have fitted the experimentally obtained data of \emph{C\textsubscript{P}}\textsubscript{-Total} to the equation (1) for \emph{T} \textgreater{}\textgreater{}
\emph{T}\textsubscript{C} where the effect of magnetic contribution is expected to be negligible \cite{gopal1966specific}. Such fitting analysis yields $\Theta$\textsubscript{D} = 488.7 K for
Ti\textsubscript{0.6}Mn\textsubscript{0.4}Co\textsubscript{2}O\textsubscript{4} polycrystalline sample which is 37 K lower than the $\Theta$\textsubscript{D} of antiferromagnetic Co\textsubscript{3}O\textsubscript{4} and 72 K lower than the $\Theta$\textsubscript{D} of ferrimagnetic
TiCo\textsubscript{2}O\textsubscript{4} \cite{roth1964magnetic,ogawa1965specific}. The extrapolate data
is shown in figure 11 together with the contribution of the \emph{C\textsubscript{P}}\textsubscript{-Lattice} and \emph{C\textsubscript{P}}\textsubscript{-Magnetic} components obtained from the above analysis \cite{bouvier1991specific}. For \emph{T} \textless{} 47.8 K the contribution from \emph{C\textsubscript{P}}\textsubscript{-Magnetic} starts dominating over \emph{C\textsubscript{P}}\textsubscript{-Lattice} due to the freezing down of phonons and the reorientation of spins in
preferred directions. It is interesting to note that \emph{C\textsubscript{P}}\textsubscript{-Magnetic} remains constant (13 J/mol K) between 62 and 108 K. The absence of a sharp peak in the
\emph{C}\textsubscript{P}(\emph{T}) motivated us to perform the dynamic response of the magnetic susceptibility and its temperature and frequency dependent characteristics with the aim to probe the degree of disorderness and the existence of spin-glass nature in these systems. Figure 12 shows the temperature dependence of real and imaginary components of the ac-magnetic susceptibility
$\chi$\textsubscript{ac}(\emph{f},\emph{T}) = $\chi$\textsuperscript{$\prime$}(\emph{f},\emph{T}) + \emph{i} $\chi$\textsuperscript{$\prime\prime$}( \emph{f},\emph{T}) recorded at constant driving
frequency 51 Hz and peak-to-peak amplitude of the ac-magnetic field \emph{h}\textsubscript{ac} $\sim$ 4 Oe for different dc-bias-fields (0 $\leq$ \emph{H}\textsubscript{dc} $\leq$ 300 Oe) superimposed
with \emph{h}\textsubscript{ac}. Both the in-phase and out-of-phase susceptibility curves exhibit cusp like behavior with significant decreases in the peak position (\emph{T}\textsubscript{P}) and intensity
as \emph{H}\textsubscript{dc} increases. For \emph{x} = 0.4 (0.2), T\textsubscript{P} shifts from 110.7 K (78.22 K) to 102.4 K (71 K) for \emph{H}\textsubscript{dc} increases from 0 to 300 Oe. The inset of
figure 12(b) shows the variation of \emph{T}\textsubscript{P} as a function of \emph{H}\textsubscript{dc}\textsuperscript{2/3} which shows a straight--line behavior consistent with the de Almeida-Thouless line (AT-line) {[}\(H_{\text{AT}}\left( T_{P} \right) = A_{\text{AT}}\left( 1 - \frac{T_{P}(H)}{T_{P}(0)} \right)^{\alpha}\){]} \cite{de1978stability,souletie1985critical,nakamura1993spin}. Analysis, usually adopted to probe the spin-glass like magnetic disordered state in variety of compounds. Here AT line defines the onset of a transition from the frozen spin-glass like state to reversible magnetic behavior at the peak temperature in $\chi$\textsubscript{ac}(\emph{f},\emph{T}) under zero dc-field. It is interesting to note that the in-phase susceptibility cusp splits into two peaks with increasing the magnitude of \emph{H}\textsubscript{dc} (figure not shown). The extent of peak splitting increases with increasing \emph{H}\textsubscript{dc} whereas for the out of phase susceptibility $\chi$\textsuperscript{$\prime\prime$}, only a single cusp was noticed which decreases in amplitude and shifts to lower temperatures with increasing \emph{H}\textsubscript{dc}. The AT-line criteria are often used in the literature to study the characteristic features of spin-glass ordering in amorphous magnets and magnetic ultrafineparticles \cite{nakamura1993spin,mukadam2005dynamics,singh2008almeida}. The linear behavior of the plot \emph{T}\textsubscript{P} vs. \emph{H}\textsubscript{dc}\textsuperscript{2/3} for low applied fields confirms the AT-like phase boundary where the linear extrapolation of AT-line on the temperature axis, in the limit \emph{H}\textsubscript{dc} approaches to zero yields the freezing temperature
\emph{T}\textsubscript{F} (\emph{i.e}. \emph{T}\textsubscript{P}(0) $\sim$ 78.15 and 110.5 K for \emph{x} = 0.2 and 0.4, respectively). Such AT-line crossover in the \emph{H}-\emph{T} plane has
been reported for a variety of spin-glass compounds such as CuMn \cite{djurberg1999magnetic}, AgMn \cite{bouchiat1986critical}, AuFe \cite{taniguchi1988spin}, Fe$_{x}$Mn$_{1-x}$TiO\textsubscript{3} \cite{katori1993magnetic} etc. However, determination of the freezing temperature using the AT-line criteria as a standard proof is not enough to probe the spin-glass behavior, instead empirical scaling laws are needed \cite{shtrikman1981theory,binder1986spin,Mydosh1993,tholence1980frequency}.

Therefore, in order to probe the existence of exact spin-glass nature in the current system we performed a detailed frequency dependent study. Figure 13 shows the temperature variation of $\chi$\textsuperscript{$\prime$} and $\chi$\textsuperscript{$\prime\prime$} measured at various frequencies between 0.17 Hz and 1.2 kHz) under heating condition without any superimposition of
\emph{H}\textsubscript{dc} with \emph{h}\textsubscript{ac} ($\sim$ 4 Oe) for both the compositions. One can clearly notice a shift in the maximum point of the cusp towards high temperature side with increasing the frequency for both the systems, however, for \emph{x} = 0.4 this shift is more significant than the lower composition \emph{x} = 0.2. Usually, such dispersion in the peak position signifies the spin-disorderness (glassyness) in the system. In order to confirm the existence of such spin-randomness (spin-glass nature) we used two empirical scaling laws: (\emph{i}) the Vogel--Fulcher-law (VFL)$\tau$ = $\tau$\textsubscript{0} exp (\emph{E}\textsubscript{a}/k\textsubscript{B}
(\emph{T\textsubscript{f}}$-$\emph{T}\textsubscript{0}) for interacting particle systems, and (\emph{ii}) the Power-law (PL)$\tau$ = $\tau$\textsubscript{0} {[}(\emph{T}/\emph{T}\textsubscript{F})
$-$1)\textsuperscript{$-$z$\nu$}{]} describing the critical slowing down \cite{Mydosh1993,tholence1980frequency}. In the above expressions, the parameter $\tau$\textsubscript{0} represents the relaxation time constant, \emph{T}\textsubscript{0} is a measure of the interparticle interaction strength on magnetic relaxation, \emph{k}\textsubscript{B} is the Boltzmann constant, \emph{E}\textsubscript{a} is an activation energy parameter, z$\nu$ is a critical exponent and \emph{T}\textsubscript{F} is the spin-glass freezing temperature. Figure 13(e) shows the logarithmic variation of the relaxation time ($\tau$) as a function of ln (\emph{T} $-$
\emph{T}\textsubscript{F})/\emph{T}\textsubscript{F}. The scattered points represent the variation of the peak temperature (\emph{T}\textsubscript{P}) determined from $\chi$\textsuperscript{$\prime$}(\emph{T}) for both the compositions. The solid continuous lines connecting the experimental data points represent the best fit to the PL. On the other hand, figure 13(f) shows the variation of 1/ln(\(\tau/\tau_{0})\) versus \emph{T}\textsubscript{P} related to the VFL described above. Here \(\tau_{0}\) values obtained from the PL have been used to fit the experimental data points for VFL. In both the
cases (VFL and PL) we obtained the straight line behavior with different fitting parameters which are listed in table 3. The following fitting parameters are evaluated from PL for \emph{x} = 0.2(0.4): \(\tau_{0}\) = 1.48$\times$10\textsuperscript{-14} s (1.6$\times$10\textsuperscript{-15} s),
\emph{T}\textsubscript{F} = 78.1 K (110.32 K) and z\(\nu\) = 4.22 (5.23) using the frequency variation of $\chi$\textsuperscript{$\prime$}(\emph{T}). Alternatively, using the VFL we obtained for \emph{x} = 0.2(0.4): \emph{E}\textsubscript{a} = 10.3 k\textsubscript{B} (32.02 k\textsubscript{B}), and
\emph{T}\textsubscript{0} = 77.82 K (109.56 K). Using the fitting parameters obtained from VFL and the relation between the VFL and PL (i.e.\(\ \ln\left( \frac{40k_{B}T_{0}}{E_{a}} \right)\sim\frac{25}{\text{z$\nu$}}\) ) \cite{souletie1985critical}, we have reevaluated the magnitudes of \(\text{z$\nu$\ }\)which are in excellent agreement with the values determined from PL (table 3). In general, typical spin-glass system exhibit `$\tau$\textsubscript{o}' values between 10\textsuperscript{-9} and 10\textsuperscript{-16} s and the z$\nu$ values between 4 and 12 \cite{Mydosh1993,souletie1985critical}. The $\tau$\textsubscript{o} and z$\nu$ values for
Ti\textsubscript{0.6}Mn\textsubscript{0.4}Co\textsubscript{2}O\textsubscript{4} and
Ti\textsubscript{0.8}Mn\textsubscript{0.2}Co\textsubscript{2}O\textsubscript{4} lies within the above specified range for spin glasses. Thus the investigated compounds exhibit a spin-glass state just below the ferrimagnetic ordering due to the Mn substitution in TiCo\textsubscript{2}O\textsubscript{4} (often called as reentrant spin-glass state).

\begin{figure}[t]
\includegraphics[trim=0cm 0cm 0cm 0.3cm, clip=true,scale=0.3]{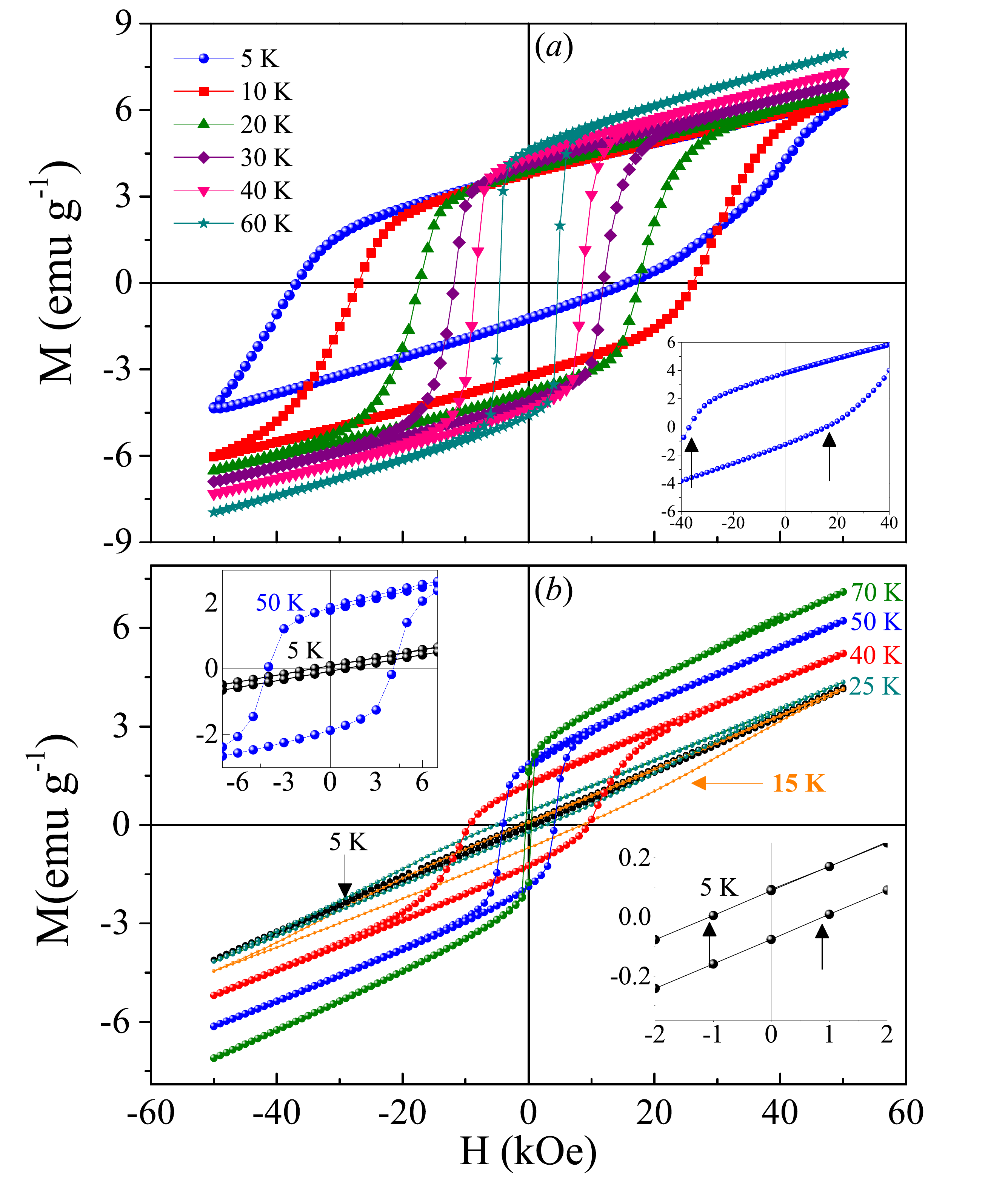}
\caption{Magnetization vs. field (\emph{M}-\emph{H}) hysteresis loops recorded at different temperatures after field cooling for ($a$) Ti\textsubscript{0.6}Mn\textsubscript{0.4}Co\textsubscript{2}O\textsubscript{4}, and ($b$) Ti\textsubscript{0.8}Mn\textsubscript{0.2}Co\textsubscript{2}O\textsubscript{4}. The insets show an enlarged view of the \emph{M}-\emph{H} loops measured at 5 K.}
\label{fig:Fig14}
\end{figure}

\subsection{\label{sec:level5}Magnetic Hysteresis and Remanence:}
Figure 14 shows the magnetization versus field (\emph{M}-\emph{H}) hysteresis loops of  Ti\textsubscript{0.6}Mn\textsubscript{0.4}Co\textsubscript{2}O\textsubscript{4}, and Ti\textsubscript{0.8}Mn\textsubscript{0.2}Co\textsubscript{2}O\textsubscript{4} measured between$-$50 kOe to +50 kOe at selected temperatures below \emph{T}\textsubscript{C} after 50 kOe field cooling. The temperature of the sample was raised to 300 K (\textgreater{}\textgreater{} \emph{T}\textsubscript{C}) before the \emph{M}-\emph{H} measurement to ensure perfect paramagnetic state to avoid any remanence field in the samples. All the hysteresis loops are unsaturated even up to 50 kOe and the magnetization increases linearly with field. Also, the spontaneous magnetization per formula unit is slightly larger than the magnitude of magnetic moments determined from the neutron diffraction and the linear contribution represents the non-collinear arrangement of spins. The angle of the spins usually changes due to the torque exerted by the magnetic field, thus leading to a linear increase in the magnetization value with increasing the field within the magnetically ordered state. The hysteresis loops of Ti\textsubscript{0.6}Mn\textsubscript{0.4}Co\textsubscript{2}O\textsubscript{4} exhibit giant coercivity (\emph{H}\textsubscript{C} $\sim$ 26.3 kOe at 5 K) and loop asymmetry [loop-shift popularly known as exchange-bias, \emph{H}\textsubscript{EB} $\sim$ $-$10.6 kOe at 5 K)] due to very high magnetocrystalline anisotropy dominant at low temperature. The magnitude of \emph{H}\textsubscript{C} and \emph{H}\textsubscript{EB} are estimated using the relations \emph{H}\textsubscript{C} = (\emph{H}\textsuperscript{+} $-$ \emph{H}$^{-}$)/2, and \emph{H}\textsubscript{EB} = (\emph{H}\textsuperscript{+} + \emph{H}$^{-}$)/2, where the quantities \emph{H}\textsuperscript{+} and \emph{H}$^{-}$ are the magnetic field values at which magnetization becomes zero. Both \emph{H}\textsubscript{C} and \emph{H}\textsubscript{EB}  decrease gradually as the temperature approaches \emph{T}\textsubscript{C}. We also observed that a weak ferrimagnetic component \emph{M}\textsubscript{WFi} is essentially superimposed on a linear AF component in the ordered state. The temperature variation of both \emph{H}\textsubscript{C} and \emph{H}\textsubscript{EB} are shown in figure 15 (i) and (iii) for the compositions \emph{x} = 0.2 and 0.4, respectively. At \emph{T} = 5 K, the \emph{H}\textsubscript{C} value of Ti\textsubscript{0.6}Mn\textsubscript{0.4}Co\textsubscript{2}O\textsubscript{4} reaches a maximum of 26.3 kOe which gradually decreases to 4.6 kOe at 60 K. On the other hand, \emph{H}\textsubscript{EB} reaches a minimum value of $-$10.6 kOe at 5 K and a sign reversal occurs at 9.7 K finally reaches a maximum value of +0.6 kOe at 10 K. Beyond 10 K, \emph{H}\textsubscript{EB} gradually deceases and approaches to zero. For the composition Ti\textsubscript{0.8}Mn\textsubscript{0.2}Co\textsubscript{2}O\textsubscript{4} no significant coercivity or remanence was noticed in the \emph{M}-\emph{H} loops recorded below 5 K, but as the temperature increases loops open-up and exhibits significant \emph{H}\textsubscript{C} and \emph{M}\textsubscript{R} values for \emph{T} $\textgreater$ 5 K. The magnitude of \emph{H}\textsubscript{C} gradually increases with increasing the temperature (from \emph{H}\textsubscript{C} $\sim$ 0.98 kOe at 5 K) and reaches its maximum value of 9.4 kOe at 40 K and drops to 0.44 kOe at 70 K. While a crossover from +3.7 kOe to $-$1.3 kOe was noticed in the \emph{H}\textsubscript{EB} values when the temperature increases from 5 to 25 K, respectively. Such polarity change in the exchange bias is also associated with the highly frustrated magnetic spins and negative magnetization behavior. For \emph{T} $\textgreater$ 25 K, the \emph{H}\textsubscript{EB} decreases and gradually approaches to zero as \emph{T} $\rightarrow$ \emph{T}\textsubscript{C}.  Similarly, we have evaluated the remanence magnetization \emph{M}\textsubscript{R} value at \emph{H} = 0 and the maximum magnetization \emph{M}\textsubscript{MAX} (at 50 kOe) for both the samples whose temperature dependence was shown in figures 15 (ii) and (iv). Interestingly, both \emph{M}\textsubscript{R} and \emph{M}\textsubscript{MAX} increase progressively with the rise of temperature for \emph{x} = 0.2 and 0.4. The maximum \emph{M}\textsubscript{R} = 4.58 emu/g and 1.9 emu/g was observed for \emph{x} = 0.2 and 0.4, respectively. A clear anomaly in \emph{M}\textsubscript{R}(\emph{T}) was noticed across 15 K for \emph{x} = 0.2 which is consistent with the field induced transition noticed in $\chi$\textsubscript{ZFC}(\emph{T}) (Figure 10(b)) and \emph{M}\textsubscript{R} decreases slightly across the \emph{T}\textsubscript{COMP} for this system.

\begin{figure}[t]
\includegraphics[trim=0cm 2cm 0cm 0.3cm, clip=true,scale=0.3]{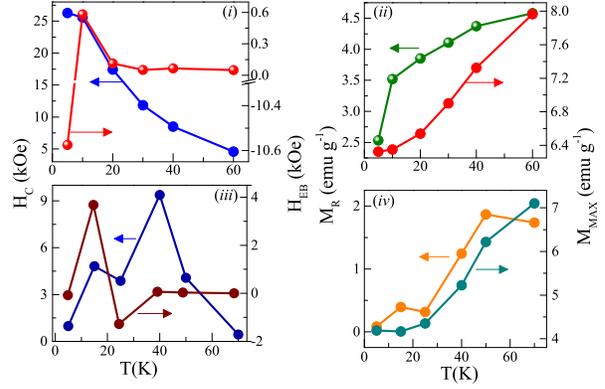}
\caption{ Temperature dependence of coercive field \emph{H}\textsubscript{C} and the exchange bias field \emph{H}\textsubscript{EB} (($i$), ($iii$)).  The remanence magnetization [\emph{M}\textsubscript{R}] and high field (50 kOe) magnetization [ \emph{M}\textsubscript{max}] (($ii$), ($iv$)) of  Ti\textsubscript{0.6}Mn\textsubscript{0.4}Co\textsubscript{2}O\textsubscript{4}, and Ti\textsubscript{0.8}Mn\textsubscript{0.2}Co\textsubscript{2}O\textsubscript{4} are also shown.}
\label{fig:Fig15}
\end{figure}

\section{\label{sec:leveIV.} Conclusions}
The structural, electronic and magnetic properties of two different compositions of a \emph{B}-site substituted Ti$_{1-x}$Mn$_{x}$Co\textsubscript{2}O\textsubscript{4} system have been investigated in detail by neutron diffraction, X-ray photoelectron spectroscopy (XPS), specific-heat, static and dynamic magnetization measurements. Based on the evidences gathered from the dc-magnetization data and ac-susceptibility we have specifically chosen two compositions \emph{x} = 0.2 and 0.4 for a detailed neutron diffraction study because these two systems exhibit evidence for the negative magnetization, magnetic compensation effect, reentrant spin-glass behavior and exchange-bias effect. The refinements of the crystal structure in the space group \emph{I}4\textsubscript{1}/\emph{amd} provides the evidence
of a weak tetragonal distortion with \emph{c}/\emph{a} \textless{} 1 for the compositions \emph{x} = 0.2 and 0.4, respectively. Due to the dimness of this distortion the presence of a Jahn-Teller effect is not really supported by these data. However, the refinements in the tetragonal structure give rise to weak changes of the apical and equatorial bond lengths. These changes are compatible with the trend which is expected for a Jahn-Teller activity of the Mn\textsuperscript{3+} ions having the 3\emph{d}\textsuperscript{4} configuration. The elemental analysis performed by XPS confirms the presence of trivalent state of all the octahedrally coordinated \emph{B}-site cations {[}\emph{T} =
Ti(3\emph{d}\textsuperscript{1}), Mn(3\emph{d}\textsuperscript{4}) and Co(3\emph{d}\textsuperscript{6}){]} and divalent electronic state of tetrahedrally coordinated \emph{A} site Co which are in-line with the neutron diffraction and magnetization data. Our neutron diffraction study clearly shows the diffuse
neutron scattering from 200 reflections which indicate the presence of canted local spin configuration (Yafet-Kittel like ordering) due to the transverse spin component at low-temperatures. Observation of the diffuse scattering and variation in canting angle with the Mn substitution are the most significant features of the current investigated system. Neutron diffraction studies also confirm the presence of a strong ferrimagnetic (FI) coupling between the magnetic \emph{A}- and \emph{B}-site ions below \emph{T}\textsubscript{C} = 110.3 K (for \emph{x} = 0.4) and 78.2 K (for \emph{x} = 0.2). Furthermore, an
additional weak antiferromagnetic (AF) component was found lying perpendicular to the FI component. Although the moment directions of AF and FI components could not be determined, but it is possible to
determine the spin sequence of the AF ordering in which the moments were assumed to be aligned parallel or perpendicular the \emph{c} axis. Thermal variation of the individual and net magnetic moments 2$\mu$\textsubscript{FI}(\emph{T\textsubscript{B}}) $-$~$\mu$\textsubscript{FI}(Co\emph{\textsubscript{A}}),
(\emph{T\textsubscript{B}} being the cations at \emph{B} sites) obtained from neutron diffraction agree more or less with the temperature dependence of the dc-magnetic susceptibility $\chi$\textsubscript{dc}(\emph{T}) data. However, the moments $\mu$\textsubscript{FI}(Co)\textsubscript{\emph{A}} and
$\mu$\textsubscript{FI}(\emph{T\textsubscript{B}}) gradually decreases to zero in consonance with the $\chi$\textsubscript{dc}(\emph{T}), but could not able to see any negative moment or any anomaly across the compensation point ($\sim$25.4 K for \emph{x} = 0.2) which were observed in the $\chi$\textsubscript{dc}(\emph{T}) data. This discrepancy is due to the fact that the neutron diffraction measurements were performed under zero-field case, whereas the dc magnetization experiments were performed in the presence of strong dc-field. The current neutron diffraction results demonstrate that the unequal growth of magnetic moments in the cations at tetrahedral \emph{A} and octahedral \emph{B} sites, and their different temperature dependence plays a significant role in realizing the magnetization reversal for all the compositions \emph{x} $\leq$ 0.2. Interestingly as compared to the parent compound TiCo\textsubscript{2}O\textsubscript{4},  Ti\textsubscript{0.6}Mn\textsubscript{0.4}Co\textsubscript{2}O\textsubscript{4} sample exhibit giant coercive field \emph{H}\textsubscript{C} ($\sim$26 kOe), and negative exchange bias \emph{H}\textsubscript{EB} ($-$10.6 kOe) at low temperatures which vanishes as the measuring temperature approaches towards \emph{T}\textsubscript{F}. From the temperature dependence of the total specific-heat \emph{C\textsubscript{P}}\textsubscript{-Total}(\emph{T}) we have evaluated the individual contributions of lattice and magnetic components. Accordingly, we have determined the Debye temperature $\Theta$\textsubscript{D} = 488.7 K (for
Ti\textsubscript{0.6}Mn\textsubscript{0.4}Co\textsubscript{2}O\textsubscript{4})which is significantly lower (by 65.5 K) than the $\Theta$\textsubscript{D} value of single crystal TiCo\textsubscript{2}O\textsubscript{4}. The frequency and temperature dependence of the dynamic-susceptibility $\chi$\emph{\textsubscript{ac}}(\emph{f},\emph{T}) data follows the AT-line criterion in \emph{H}-\emph{T} plane and empirical scaling laws (i.e. Vogel$-$Fulcher-law and Power-Law) for both the compositions \emph{x} = 0.2 and 0.4 suggesting the spin-glass like ordering below \emph{T}\textsubscript{F} = 78.1 and 110.1 K with critical exponents z$\nu$ = 4.22 and 5.23, and relaxation time constant $\tau$\textsubscript{0} = 1.48 $\times$ 10\textsuperscript{-14} and 1.6 $\times$ 10\textsuperscript{-15} s.

\section*{Acknowledgements:}
P.P. acknowledge FIST programme of Department of Science and Technology, India for partial support of this work (Ref. No: SR/FST/PSII-020/2009 and SR/FST/PSII-037/2016). T.S. acknowledges financial support from the Swedish Research Council (VR starting grant number: 2017-05030). R. S. M. acknowledges the financial support from SERB for his Early Career Research Award (File no.: ECR/2018/000999/PMS).

\section*{References}


\pagebreak

\begin{table*}[t]
\begin{center}
\caption{Results of the Rietveld refinements of the neutron powder
diffraction data of
Ti\textsubscript{0.2}Mn\textsubscript{0.8}Co\textsubscript{2}O\textsubscript{4}
and
Ti\textsubscript{0.4}Mn\textsubscript{0.6}Co\textsubscript{2}O\textsubscript{4}
collected on the instrument E9 at 3 K. The refinements were carried out
in the tetragonal space group \emph{I}4\textsubscript{1}/\emph{amd}. The
given residuals are defined as \emph{R\textsubscript{F}} =
$\sum$\textbar{}\textbar{}\emph{F}\textsubscript{obs}\textbar{} $-$
\textbar{}\emph{F}\textsubscript{calc}\textbar{}\textbar{}/$\sum$\textbar{}\emph{F}\textsubscript{obs}\textbar{}).
Listed are the positional parameters \emph{y} and \emph{z} of the O atom
located at the site 16\emph{h}(0,\emph{y},\emph{z}) and the occupancies
of the Ti\emph{\textsubscript{B}}, Mn\emph{\textsubscript{B}}, and
Co\emph{\textsubscript{B}} atoms at the site 8\emph{c}(0,0,0). For all
atoms an overall thermal parameter \emph{B}\textsubscript{iso} was
refined. Further the bond distances in the
\emph{T\textsubscript{B}}O\textsubscript{6} octahedra
(\emph{T\textsubscript{B}} = Ti\emph{\textsubscript{B}},
Mn\emph{\textsubscript{B}}, and Co\emph{\textsubscript{B}}), and the
Co\emph{\textsubscript{A}}O\textsubscript{4} tetrahedra as well as the
lattice parameters are also given.}

\vspace{0.3cm}
\setlength{\arrayrulewidth}{0.2mm}
\setlength{\tabcolsep}{30pt}
\renewcommand{\arraystretch}{1.5}
\begin{tabular}{|cccc|}
\hline
 & Ti\textsubscript{0.8}Mn\textsubscript{0.2}Co\textsubscript{2}O\textsubscript{4}  &  Ti\textsubscript{0.6}Mn\textsubscript{0.4}Co\textsubscript{2}O\textsubscript{4} &  \\
\hline
\emph{T}[K] & 3 & 3 & 295 \\
\emph{occ}(Ti$_{B}$) & 0.835(4) & 0.618(3) & 0.628(6) \\
\emph{occ}(Mn$_{B}$) & 0.209(4) & 0.412(3) & 0.419(6) \\
\emph{occ}(Co$_{B}$) & 0.956(4) & 0.970(3) & 0.953(6) \\
\emph{y}(O) & 0.5218(5) & 0.5241(4) & 0.5246(4) \\
\emph{z}(O) & 0.2398(5) & 0.2414(5) & 0.2418(5) \\
\emph{a}\textsubscript{t}[\AA] & 5.9499(4) & 5.9231(4) & 5.9331(5) \\
\emph{c}\textsubscript{t}[\AA] & 8.4044(10) & 8.3677(11) & 8.3867(13) \\
\emph{c}\textsubscript{t}/\emph{a}\textsubscript{t}$\sqrt{2}$ & 0.9988(3) & 0.9988(3) & 0.9995(3) \\
\emph{V}[\AA$^{3}$] & 297.53(5) & 293.57(6) & 295.22(7) \\
\emph{d}$_{ab}$(\emph{T}$_{B}$-O)$\times$4 & 2.016(2) & 1.998(2) & 1.997(2) \\
\emph{d}$_{c}$(\emph{T}$_{B}$-O)$\times$2 & 2.020(5) & 2.025(4) & 2.033(4) \\
\emph{d}(Co$_{A}$-O)$\times$4 & 1.976(4) & 1.972(3) & 1.976(3) \\
\emph{R}$_{F}$ & 0.042 & 0.042 & 0.041 \\
\hline
\end{tabular}
\label{tab:table1}
\end{center}
\end{table*}

\begin{table*}[b]
\begin{center}
\caption{Magnetic moments of the transition metal atoms in
Ti\textsubscript{0.2}Mn\textsubscript{0.8}Co\textsubscript{2}O\textsubscript{4}
and
Ti\textsubscript{0.4}Mn\textsubscript{0.6}Co\textsubscript{2}O\textsubscript{4}
obtained from Rietveld refinements using neutron diffraction data
collected on the instruments E2 and E6 at 2 K, respectively. The magnetic
Co\textsuperscript{3+}, Ti\textsuperscript{3+} and
Mn\textsuperscript{3+} ions (\emph{B} site) in the space group are
located at the positions: (1) 0,0,0; (2) $\sfrac{3}{4}$, $\sfrac{1}{4}$, $\sfrac{1}{2}$; (3) $\sfrac{1}{4}$, $\sfrac{1}{2}$, $\sfrac{3}{4}$; (4) $\sfrac{1}{2}$, $\sfrac{3}{4}$,$\sfrac{1}{4}$.
The Co\textsuperscript{2+} ions (\emph{A} site) are located at (1)
$\sfrac{3}{8}$, $\sfrac{3}{8}$, $\sfrac{3}{8}$; (2) $\sfrac{1}{8}$, $\sfrac{5}{8}$, $\sfrac{1}{8}$. During the refinements we have used the constraint
$\mu$(Co\textsuperscript{3+}) = $\mu$(Mn\textsuperscript{3+}) = 4 $\times$
$\mu$(Ti\textsuperscript{3+}), and in the lower part
$\mu$(Mn\textsuperscript{3+}) = 4 $\times$ $\mu$(Ti\textsuperscript{3+}) to estimate
the moment of Co\textsuperscript{3+}. Here the moment of
Mn\textsuperscript{3+} was determined earlier for
MnCo\textsubscript{2}O\textsubscript{4} \cite{boucher1970etude}. The averaged moments
at the \emph{T\textsubscript{B}} site are also listed. The resultant
ferromagnetic moments are compared with spontaneous magnetizations
\emph{M} measured at 5 K.}
\vspace{0.3cm}
\setlength{\arrayrulewidth}{0.2mm}
\setlength{\tabcolsep}{12pt}
\renewcommand{\arraystretch}{1.5}
\begin{tabular}{|ccccc|}
\hline
 Magnetic moment & Ti\textsubscript{0.8}Mn\textsubscript{0.2}Co\textsubscript{2}O\textsubscript{4}  &  Ti\textsubscript{0.8}Mn\textsubscript{0.2}Co\textsubscript{2}O\textsubscript{4} &  Ti\textsubscript{0.6}Mn\textsubscript{0.4}Co\textsubscript{2}O\textsubscript{4} & Ti\textsubscript{0.6}Mn\textsubscript{0.4}Co\textsubscript{2}O\textsubscript{4}  \\
  & E2 & E6 & E2 & E6  \\
\hline
$\mu$\textsubscript{FI}(Co$_{A}$)[$\mu$\textsubscript{B}] & 2.60(3) & 2.99(2) & 2.94(4) & 3.10(3)  \\
$\mu$\textsubscript{FI}(Mn$_{B}$/Co$_{B}$)[$\mu$\textsubscript{B}] & 2.09(5) & 2.29(4) & 2.13(4) & 2.39(4) \\
$\mu$\textsubscript{FI}(Ti$_{B}$)[$\mu$\textsubscript{B}] &  0.52(1) & 0.57(1) & 0.53(1) & 0.60(1)  \\
$\mu$\textsubscript{FI}(\emph{T}$_{B}$)[$\mu$\textsubscript{B}] & 1.43(3) & 1.57(3) & 1.63(3) & 1.82(3)  \\
$\mu$\textsubscript{AF}(Mn$_{B}$/Co$_{B}$)[$\mu$\textsubscript{B}] & 0.90(4) & 0.68(4) & 0.72(3) & 0.68(3) \\
$\mu$\textsubscript{AF}(Ti$_{B}$)[$\mu$\textsubscript{B}] & 0.22(1) & 0.17(1) & 0.18(1) & 0.17(1)  \\
$\mu$\textsubscript{AF}(\emph{T}$_{B}$)[$\mu$\textsubscript{B}] & 0.61(2) & 0.46(2) & 0.56(2) & 0.52(2)  \\
$\mu$\textsubscript{tot}(Mn$_{B}$/Co$_{B}$)[$\mu$\textsubscript{B}] & 2.27(4) & 2.39(3) & 2.25(3) & 2.48(3) \\
$\mu$\textsubscript{tot}(Ti$_{B}$)[$\mu$\textsubscript{B}] & 0.57(1) & 0.60(1) & 0.56(1) & 0.62(1)  \\
$\mu$\textsubscript{tot}(\emph{T}$_{B}$)[$\mu$\textsubscript{B}] &1.55(2) & 1.64(2) & 1.72(2) & 1.90(2)  \\
\emph{R\textsubscript{F}/R}\textsubscript{M} & 0.019 / 0.041 & 0.007 / 0.042 & 0.023 / 0.041 & 0.012 / 0.052  \\
2$\mu$\textsubscript{FI}(\emph{T}$_{B}$)$-$$\mu$\textsubscript{FI}(Co$_{A}$)[$\mu$\textsubscript{B}] & 0.25(10) & 0.15(8) & 0.31(8) & 0.55(7) \\
\emph{M} {[}$\mu$\textsubscript{B}{]} & 0.32 & 0.32 & 0.35 & 0.35 \\
\hline
$\mu$\textsubscript{tot}(Ti\emph{\textsubscript{B}}) {[}$\mu$\textsubscript{B}{]} & 0.96 & 0.96 & 0.96 & 0.96 \\
$\mu$\textsubscript{tot}(Mn\emph{\textsubscript{B}}) {[}$\mu$\textsubscript{B}{]} & 3.84 & 3.84 & 3.84 & 3.84 \\
$\mu$\textsubscript{tot}(Co\emph{\textsubscript{B}}) {[}$\mu$\textsubscript{B}{]} & 0.75(5) & 0.80(4) & 0.60(4) &
0.65(3) \\
\hline
\end{tabular}
\label{tab:table1}
\end{center}
\end{table*}

\begin{table*}[t]
\begin{center}
\caption{The list of various fitting parameters obtained from straight line fitting of ac-susceptibility data using the Vogal-Fulcher law (VFL) and Power law (PL).}
\vspace{0.3cm}
\setlength{\arrayrulewidth}{0.1mm}
\setlength{\tabcolsep}{4pt}
\renewcommand{\arraystretch}{1.3}
\begin{tabular}{|cccc|}
\hline
 Sample & Fitted parameters corresponding to PL & Fitted parameters corresponding to VFL & z$\nu$ values from  \\
  & $\tau$ = $\tau$\textsubscript{0}
((\emph{T}/\emph{T}\textsubscript{F}) $-$1)\textsuperscript{-z$\nu$} &  $\tau$ =
$\tau$\textsubscript{0} exp($\frac{\emph{E}\textsubscript{a}}
{k\textsubscript{B}(\emph{T}_{f} - \emph{T}\textsubscript{0})}$) &  ln⁡($\frac{40k\textsubscript{B}\emph{T}\textsubscript{0}}{\emph{E}\textsubscript{a}}$) $\sim$ $\frac{25}{z\nu}$ \\
\hline
 & Using $\chi$\textsuperscript{$\prime$}($f$,T) & Using $\chi$\textsuperscript{$\prime$}($f$,T) &  \\
 & $\tau$\textsubscript{0} = 1.48$\times$10\textsuperscript{-14} s & $\tau$\textsubscript{0} = 1.48$\times$10\textsuperscript{-14} s & \\
  & \emph{T}\textsubscript{F} = 78.1$\pm$0.02 K & \emph{E}\textsubscript{a} = 10.3 k\textsubscript{B} & 4.37  \\
   & z$\nu$= 4.22$\pm$0.01 & \emph{T}\textsubscript{0}= 77.82$\pm$0.02 K &  \\
Ti$_{0.8}$Mn$_{0.2}$Co$_{2}$O$_{4}$ &  & &  \\
& Using $\chi$\textsuperscript{$\prime\prime$}($f$,T) & Using $\chi$\textsuperscript{$\prime\prime$}($f$,T) &  \\
 & $\tau$\textsubscript{0} = 3.67$\times$10\textsuperscript{-15} s & $\tau$\textsubscript{0} = 3.67$\times$10\textsuperscript{-15} s &  \\
  & \emph{T}\textsubscript{F} = 78.03$\pm$0.02 K & \emph{E}\textsubscript{a} = 20.82 k\textsubscript{B} &  \\
   & z$\nu$= 4.6$\pm$0.02 & \emph{T}\textsubscript{0}= 77.46$\pm$0.01 K & 4.9  \\
\hline
& Using $\chi$\textsuperscript{$\prime$}($f$,T) & Using $\chi$\textsuperscript{$\prime$}($f$,T) &  \\
 & $\tau$\textsubscript{0} = 1.6$\times$10\textsuperscript{-15} s & $\tau$\textsubscript{0} = 1.6$\times$10\textsuperscript{-15} s &  \\
  & \emph{T}\textsubscript{F} = 110.1$\pm$0.02 K & \emph{E}\textsubscript{a} = 32.02 k\textsubscript{B} &  \\
   & z$\nu$= 5.23$\pm$0.01 & \emph{T}\textsubscript{0}= 109.56$\pm$0.02 K & 5.18  \\
Ti$_{0.6}$Mn$_{0.4}$Co$_{2}$O$_{4}$ &  & &  \\
& Using $\chi$\textsuperscript{$\prime\prime$}($f$,T) & Using $\chi$\textsuperscript{$\prime\prime$}($f$,T) &  \\
 & $\tau$\textsubscript{0} = 3.48$\times$10\textsuperscript{-14} s & $\tau$\textsubscript{0} = 3.48$\times$10\textsuperscript{-14} s &  \\
  & \emph{T}\textsubscript{F} = 109.99$\pm$0.02 K & \emph{E}\textsubscript{a} = 48.87 k\textsubscript{B} & \\
   & z$\nu$= 5.38$\pm$0.03 & \emph{T}\textsubscript{0}= 108.82$\pm$0.02 K & 5.94  \\
\hline
\end{tabular}
\label{tab:table1}
\end{center}
\end{table*}

\end{document}